\documentclass[prl,twocolumn, showpacs,showkeys,secnumarabic,amssymb,amsmath,superscriptaddress]{revtex4-1}
\usepackage{amssymb,amsmath,amsthm}
\usepackage{bbold}
\usepackage{color}
\usepackage{graphics}
\usepackage{graphicx}
\usepackage{enumitem}
\usepackage{bm}

\begin{document}

\title{Measurement-based deterministic  imaginary time evolution}

\author{Yuping Mao}
\affiliation{State Key Laboratory of Precision Spectroscopy, School of Physical and Material Sciences, East China Normal University, Shanghai 200062, China}
\affiliation{New York University Shanghai, 567 West Yangsi Road, Shanghai, 200126, China}

\author{Manish Chaudhary}
\affiliation{State Key Laboratory of Precision Spectroscopy, School of Physical and Material Sciences, East China Normal University, Shanghai 200062, China}
\affiliation{New York University Shanghai, 567 West Yangsi Road, Shanghai, 200126, China}

\author{Manikandan Kondappan}
\affiliation{State Key Laboratory of Precision Spectroscopy, School of Physical and Material Sciences, East China Normal University, Shanghai 200062, China}
\affiliation{New York University Shanghai, 567 West Yangsi Road, Shanghai, 200126, China}

\author{Junheng Shi}
\affiliation{CAS Key Laboratory of Theoretical Physics and Institute of Theoretical Physics, Chinese Academy of Sciences, Beijing 100190, China}
\affiliation{New York University Shanghai, 567 West Yangsi Road, Shanghai, 200126, China}

\author{Ebubechukwu O. Ilo-Okeke}
\affiliation{New York University Shanghai, 567 West Yangsi Road, Shanghai, 200126, China}
\affiliation{NYU-ECNU Institute of Physics at NYU Shanghai, 3663 Zhongshan Road North, Shanghai 200062, China}

\author{Valentin Ivannikov}
\affiliation{New York University Shanghai, 567 West Yangsi Road, Shanghai, 200126, China}
\affiliation{NYU-ECNU Institute of Physics at NYU Shanghai, 3663 Zhongshan Road North, Shanghai 200062, China}

\author{Tim Byrnes}
\email{tim.byrnes@nyu.edu}
\affiliation{New York University Shanghai, 567 West Yangsi Road, Shanghai, 200126, China}
\affiliation{State Key Laboratory of Precision Spectroscopy, School of Physical and Material Sciences, East China Normal University, Shanghai 200062, China}
\affiliation{NYU-ECNU Institute of Physics at NYU Shanghai, 3663 Zhongshan Road North, Shanghai 200062, China}
\affiliation{Shanghai Frontiers Science Center of Artificial Intelligence and Deep Learning, 567 West Yangsi Road, Shanghai, 200126, China}
\affiliation{Center for Quantum and Topological Systems (CQTS), NYUAD Research Institute, New York University Abu Dhabi, UAE}
\affiliation{Department of Physics, New York University, New York, NY, 10003, USA}

\begin{abstract}
We introduce a method to perform imaginary time evolution in a controllable quantum system using  measurements and conditional unitary operations. By performing a sequence of weak measurements based on the desired Hamiltonian constructed by a Suzuki-Trotter decomposition, an evolution approximating imaginary time evolution can be realized. The randomness due to measurement is corrected using conditional unitary operations, making the evolution deterministic. Both the measurements required for the algorithm and the conditional unitary operations can be constructed efficiently. We show that the algorithm converges only below a specified energy threshold and the complexity is estimated for some specific problem instances.  
\end{abstract}

\maketitle

\paragraph{Introduction} 

Imaginary time evolution is an important and enduring concept in several areas of quantum physics, despite not being directly a physical process \cite{sakurai1995modern}. 
In imaginary time evolution (ITE) of a quantum system with Hamiltonian $H$,  time $ t $ is replaced by imaginary time $t \rightarrow -i\tau$, such that the evolution operator is $ e^{-H \tau}$ \cite{magnus1954exponential,vidal2007classical}.
As such, for long evolution times, the state approaches the ground state of the Hamiltonian \cite{lin2021real,schuch2007computational}. 
ITE can be directly applied as a numerical procedure on classical computers to obtain low-energy states \cite{mcardle2019variational,chiofalo2000ground,PhysRevE.76.036712,liu2005bohm}. It is also central in making a formal connection between a $d$-spatial dimensional quantum field theory and a $ d+1$-dimensional classical statistical mechanics system, through the Wick rotation \cite{wick1954properties,peskin2018introduction,majid1994q}. 
A variety of classical simulation methods take advantage of this connection, such as quantum Monte Carlo and its variants \cite{vesely1994computational,lester1990quantum,jarrell1992hubbard,baroni1999reptation,byrnes2004hamiltonian}. 

As a numerical procedure on a classical computer, ITE requires exponential resources that scale with the size of the Hilbert space.  If there was a way of implementing ITE on a quantum computer efficiently, this could potentially be an extremely powerful tool. 
A direct implementation of the ITE operator $ e^{-H \tau } $, assuming elementary ITE gates, would have a complexity that scales polynomially with the number of subsystems, e.g. qubits.  In comparison to the same calculation performed on a classical computer, this would give an exponential speedup. In ITE, convergence to a high fidelity state takes a timescale of the inverse energy gap.  In a quantum simulation scenario, one is often interested in obtaining low-energy eigenstates of various systems, applicable to condensed matter physics, high-energy physics, and quantum chemistry \cite{feynman1982simulating,buluta2009quantum,byrnes2007quantum,georgescu2014quantum,cirac2012goals,gerritsma2010quantum,o2016scalable,horikiri2016high,houck2012chip,byrnes2021quantum}.    More generally, it may also be used as a general optimization tool, where a cost function is minimized \cite{mohseni2022ising}. Applied to the context of solving the generalized Ising model, a problem that can be mapped to any optimization problem in the complexity class NP in polynomial time, the approach could be used to optimize problems in a variety of contexts such as logistics, financial applications, artificial intelligence, pharmaceutical and material development \cite{lucas2014ising, tanahashi2019application, smelyanskiy2012near,hauke2020perspectives,mohseni2022ising}. Another application of ITE is as a state preparation protocol.  For applications such as quantum metrology \cite{giovannetti2011advances,toth2014quantum,you2017multiparameter} and alternative model of quantum computation \cite{raussendorf2001one,nayak2008non,abdelrahman2014coherent}, resource states need to be generated, which are sometimes difficult to produce.  By engineering a suitable Hamiltonian where the desired state is the ground state, ITE can be used to generate and stabilize the state \cite{tame2006natural,bartlett2006simple,van2008graph,kyaw2014measurement}.

Several methods have been proposed to perform ITE in a controllable quantum system. In  Variational Imaginary Time Evolution (VITE) \cite{mcardle2019variational}, McArdle,Yuan and co-workers introduced a  hybrid quantum-classical approach to achieve ITE. Here, the Schr\"{o}dinger equation is first solved in imaginary time on a classical computer to determine the parameters of a trial state, then this is used as the approximation of the quantum state for the quantum circuit. This method has been used to simulate the spectra of Hamiltonian \cite{jones2019variational}, perform generalized time evolution \cite{endo2020variational}, and to solve quantum many-body problems \cite{yuan2019theory}.
Motta, Chan and co-workers proposed the Quantum Imaginary Time Evolution (QITE) method \cite{motta2020determining}, where non-unitary time evolution is approximated by a unitary operator which contains the variation of the quantum systems \cite{yeter2020practical,gomes2020efficient,tan2020fast,kamakari2022digital,cao2022quantum}.  
This method has been applied to the study of quantum simulation \cite{nishi2021implementation}, nuclear energy level computation \cite{yeter2020practical}, and quantum chemistry \cite{gomes2020efficient}. 
In another approach, Williams proposed a probabilistic approach to  non-unitary quantum computing \cite{williams2004probabilistic}.  For example, in Probabilistic Imaginary Time Evolution (PITE) \cite{liu2021probabilistic}, an $ L $ qubit non-unitary gate simulation can be probabilistically obtained by designing an $L+1$ qubit system and measuring the ancilla qubit \cite{gingrich2004non}. When measuring the ancilla qubit, the $ L $-qubit state will collapse into the desired state with a certain probability. PITE exploits Grover's algorithm \cite{PhysRevLett.79.325} to enhance the probability of getting the desired state while maintaining a high fidelity. PITE is suggested to be applicable to quantum chemistry problems \citep{kosugi2021probabilistic}. The above ITE methods can be applied to various quantum algorithms. It has been shown that VITE can be applied to variational quantum algorithms for Boltzmann machine learning \cite{shingu2021boltzmann}, while QITE can be applied to the QLanczos algorithm \cite{motta2020determining,yeter2021scattering} and  variational quantum algorithms for Hamiltonian
diagonalization \cite{zeng2021variational}.

In this paper, we propose a general method of performing ITE in a controllable quantum system.  Our method relies upon performing measurements that mimic the ITE operator for small times.  By performing repeated measurements on the system using these measurement operators, combined with a unitary correction step that acts conditionally on the measurement outcomes, this allows for a way to drive the state towards the lowest energy state of the given Hamiltonian.  Much like quantum feedforward approaches such as in quantum teleportation, this converts the stochastic evolution into a deterministic one, such that the desired state is obtained with unit probability for sufficiently long evolution times  \cite{knill2001scheme,steffen2013deterministic,ma2012quantum}. 
The basic idea of the approach is to perform a weak measurement in the energy eigenbasis of a given Hamiltonian. During the slow collapse of the state, if the energy estimate is higher than a given threshold, then a conditional unitary is applied to disturb the system. This is repeated until the energy is sufficiently low, after which full collapse to the ground state occurs. Similar approaches were used for quantum state preparation \cite{ilo2018remote} using weak measurements \cite{ilo2014theory,ilo2016information}. Our approach differs from related works such as Refs. \cite{liu2021probabilistic}, where the desired outcome is obtained by postselection. It also differs from approaches such as in Refs.  \cite{mcardle2019variational,motta2020determining} since the use of measurements involves an explicitly non-unitary step. As such, no precomputation needs to be performed to determine the evolution path.  

\paragraph{Weak energy measurements}

We start by describing the general approach to performing ITE, then illustrate our approach with several examples.  Our aim will be to perform ITE of an arbitrary Hamiltonian $  H $, such that we obtain the ground state
\begin{align}
 e^{-H \tau} |\psi_0 \rangle \xrightarrow{\tau \rightarrow \infty} |E_0 \rangle ,
\label{imagtime}
\end{align}
where $|\psi_0 \rangle$ is an arbitrary initial state and $ |E_0 \rangle $ is the ground state of $ H $. We start by constructing measurement operators that take a similar form to the exponentiated Hamiltonian (\ref{imagtime}).  This can be achieved by performing a weak measurement of the Hamiltonian, with measurement operators 
\begin{align}
M_0 & = \langle 0 |_a e^{-i \epsilon H \otimes Y} | + \rangle_a = \frac{1}{\sqrt{2}} (\cos \epsilon H - \sin \epsilon H ) \nonumber \\
& = \frac{1}{\sqrt{2}} \sum_n ( \cos \epsilon E_n - \sin  \epsilon E_n) | E_n \rangle \langle E_n | \approx \frac{e^{-\epsilon H }}{\sqrt{2}} \label{measexp0}
\\
M_1 & = \langle 1 |_a e^{-i \epsilon H \otimes Y} | + \rangle_a =  \frac{1}{\sqrt{2}} (\cos \epsilon H + \sin \epsilon H ) \nonumber \\
& = \frac{1}{\sqrt{2}} \sum_n ( \cos \epsilon E_n + \sin  \epsilon E_n) | E_n \rangle \langle E_n | \approx \frac{e^{\epsilon H }}{\sqrt{2}} .
\label{measexp}
\end{align}
where Pauli spin operators are denoted $ X, Y, Z $,  and the approximation is valid for $ ||\epsilon  H|| \ll 1 $.    The Hamiltonian is taken to have a suitable energy offset and $ \epsilon $ is chosen such that the energy spectrum fits in the region $ -\pi/4 \le \epsilon E_n \le \pi/4 $. This measurement can be realized by preparing an ancilla qubit in the state  $ | + \rangle_a = (|0 \rangle_a + |1 \rangle_a)/\sqrt{2} $ and performing an interaction with Hamiltonian $  H \otimes Y $, and measuring the ancilla in the $ Z $-basis.  The measurement operators satisfy $ M_0^\dagger M_0 + M_1^\dagger M_1 = I $, where $I$ is the identity matrix. In the case that the interaction $ H \otimes Y $ is not directly accessible due to the Hamiltonian being composed of a sum of terms 
$ H =\sum_{j=1}^N H^{(j)} $, a Suzuki-Trotter decomposition \cite{suzuki1993improved,kapit2012non} of  $  e^{-i \epsilon H \otimes Y} $  to suitable order is instead performed  (see Supplementary Information).  This has the effect of changing the precise form of (\ref{measexp}), but is still an approximation to the imaginary time exponentiated Hamiltonian.

\begin{figure}[t]
\includegraphics[width=\linewidth]{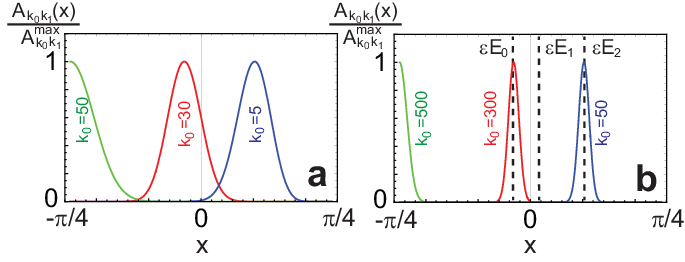}
\caption{The amplitude modulation function $ A_{k_0 k_1}(x) $ as defined in (\ref{afuncdef}).  The functions (solid lines) are normalized to their peak values, defined by 
 $ A_{k_0 k_1}^{\max} = A_{k_0 k_1}(x^{\max}_{k_0 k_1} ) $.  The total number of measurements is fixed to 
(a)  $ k_0 + k_1 = 50 $ and (b) $ k_0 + k_1 = 500 $ and the value of $ k_0 $ is as marked.   
Dashed vertical lines are values of the energy eigenstates multiplied by $ \epsilon $.}
\label{fig1} 
\end{figure}

We wish to perform the ITE to amplify the ground state as in (\ref{imagtime}). If it were possible to apply $ M_0 $ only, this would achieve a similar evolution to (\ref{imagtime}) since $ \cos \epsilon E_n - \sin \epsilon E_n $ is monotonically decreasing in the domain $ -\pi/4 \le \epsilon E_n \le \pi/4 $, such that $ M_0^k |\psi_0 \rangle \xrightarrow{k \rightarrow \infty} |E_0 \rangle $.  However, since the two outcomes $\{ M_0, M_1 \} $ occur randomly  according to quantum measurement probabilities, such a sequence is typically a rare occurrence.   Let us analyze a particular measurement sequence where there are $k_0 $ counts of $M_0 $ and $ k_1 $ counts of $ M_1 $. Since $ [M_0 , M_1 ] = 0 $, the order of the outcomes does not matter and this measurement sequence can be written
\begin{align}
M_0^{k_0} M_1^{k_1} &  | \psi_0 \rangle =  \sum_n A_{k_0 k_1}(\epsilon E_n) \langle E_n | \psi_0 \rangle | E_n \rangle
\label{postmeas}
\end{align}
where we defined the amplitude function
\begin{align}
A_{k_0 k_1}(x) & = \frac{1}{\sqrt{2^{k_0+k_1}}} ( \cos x - \sin x)^{k_0} ( \cos x + \sin x)^{k_1} \nonumber \\
& = \cos^{k_0} (x + \pi/4) \sin^{k_1} (x + \pi/4) . 
\label{afuncdef}
\end{align}
In Fig. \ref{fig1}(a) we show a plot of the function $ A $. We see that for $ -\pi/4 \le x \le \pi/4 $ and a large number of measurements it has a Gaussian form \cite{ochoa2018simultaneous}, where the peak value occurs at 
\begin{align}
x^{\max}_{k_0 k_1} = \epsilon E^{\max}_{k_0 k_1} = \frac{1}{2} \arcsin \left( \frac{k_1-k_0}{k_0+k_1} \right)
\label{gaussmax}
\end{align}
and the width is $ \sigma \approx 1/\sqrt{2(k_0 + k_1)} $.  Here, $ E^{\max}_{k_0 k_1} $ is the peak value in terms of energy.  As the number of measurements are increased, the Gaussians become increasingly well-defined (Fig. \ref{fig1}(b)).  In the limit of a large number of measurements, a collapse on the energy basis occurs.

In order to increase the amplitude of the ground state in (\ref{postmeas}), we require that 
the Gaussian is peaked with an 
outcome with $ x^{\max}_{k_0 k_1} < \epsilon (E_0+E_1)/2  $ (see Fig. \ref{fig1}(b)). 
This will create an amplitude gain of the ground state over all the remaining states, since the peak of Gaussian is closer to $\epsilon E_0$ than any other eigenvalue,  and the tail of the Gaussian on the higher energy side will suppress all higher energy states. So our strategy will then be to control the position of the Gaussian such that it lies in the desired energy range. 

\paragraph{ The algorithm}

To this end, we turn to an adaptive strategy, where a unitary operation is applied conditioned on the measurement outcomes.  Our basic strategy will be to continually monitor the location of the Gaussian using the expression (\ref{gaussmax}). 
 If the location of Gaussian corresponds to a sufficiently low energy state, then no unitary is applied.  If the Gaussian is located at a value that is of a higher energy than a chosen energy threshold 
$ E_\text{th} $, then a corrective unitary is applied.  Concretely, we iteratively perform 
\begin{align}
|\psi_{t+1} \rangle = \frac{U_{k_0^{(t+1)} k_1^{(t+1)} } M_n |\psi_{t} \rangle }{ \sqrt{\langle \psi_{t} | M_n^\dagger M_n | \psi_{t} \rangle } }
\label{generalevolution}
\end{align}
where $ n \in \{ 0, 1 \} $ labels the $(t+1)$th measurement outcome,  with
\begin{align}
U_{k_0 k_1} = \left\{
\begin{array}{ll}
I & \text{if } x^{\max}_{k_0 k_1} < \epsilon E_\text{th} \\
U_C &  \text{otherwise}
\end{array}
\right.  ,
\label{udef}
\end{align}
and 
\begin{align}
k_m^{(t+1)} = \left\{
\begin{array}{ll}
k_m^{(t)}+ \delta_{mn} & \text{if } x^{\max}_{(k_0^{(t)}+ \delta_{0n}) ( k_1^{(t)}+ \delta_{1n})  } < \epsilon E_\text{th} \\
0 & \text{otherwise}
\end{array}
\right. 
\end{align}
are the cumulative measurement outcomes starting with $ k_m^{(0)} = 0 $. In words, this counts the number of $ M_0, M_1 $ measurements respectively,  until it is found that the energy estimate is above the threshold, at which point the counts are reset to zero.  To ensure convergence of the sequence to the ground state, we demand a non-zero transition amplitude between all energy eigenstates $ |\langle E_n | U_C | E_m \rangle | > 0, \forall n,m $. For $ E_0 < E_\text{th} < E_1 $, this ensures that only the ground state is the unique fixed point of the evolution (see Supplementary Material). The requirement $ |\langle E_n | U_C | E_m \rangle | > 0 $  is not usually very difficult to satisfy since it merely requires off-diagonal matrix elements in the energy basis, which occurs for a large number of matrices. Practically, one may choose a random unitary matrix based on readily available gates.  In this way, the wavefunction for the ground state does not need to be known for the procedure.  We note that if there is some knowledge of the eigenstates $ | E_n \rangle $, then more sophisticated strategies beyond the above requirement and (\ref{udef}) can be used to construct $ U_C $.  For instance, rotations targeting the ground state based on the energy estimate  $ E^{\max}_{k_0 k_1} $  could be implemented.

\paragraph{Example 1: One qubit}

\begin{figure}[t]
\includegraphics[width=\linewidth]{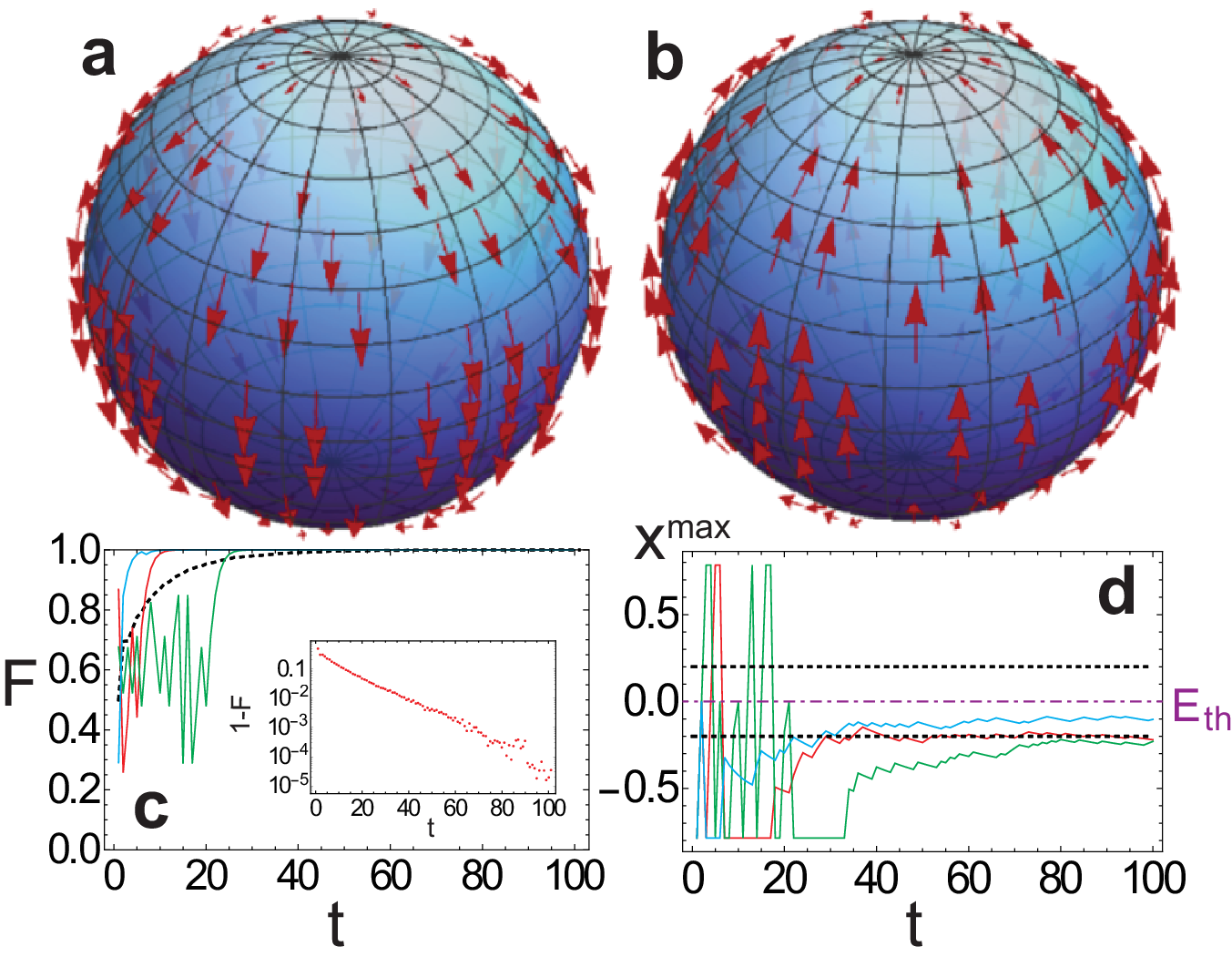}
\caption{(a)(b) Vector map on the Bloch sphere for the change induced by the operators $ M_0 $ and $ M_1 $ respectively, where $ M_n = (I \cos \epsilon  - (-1)^n Z \sin \epsilon)/\sqrt{2} $.  (c) Fidelity of the state with respect to ground state $ |E_0 \rangle = | 1 \rangle $ for the Hamiltonian $ H = Z $ after $ t $ rounds of measurement and correction under  (\ref{generalevolution}) for three random initial states (solid lines) and $ \epsilon = 0.2 $. We take $ U_C = X $ and $ E_{\text{th}} = 0 $.  Dashed line shows the averaged fidelity of 1000 evolutions starting from the initial state $ | + \rangle $. Inset shows a semilog plot of $ 1-F $ with $ t $.  (d) The peak position $ x_{k_0 k_1}^{\max} $ as defined in (\ref{gaussmax}) of the function $ A $ (solid lines). Dashed lines show the energy eigenstates $ \epsilon E_n $ and the dashed dotted line $ E_{\text{th}} $.  For the measurements in (c)(d), the outcomes are chosen randomly according to Born probabilities. }
\label{fig2} 
\end{figure}

We start with the simplest example of a single qubit with Hamiltonian $ H = Z  $.   In Fig. \ref{fig2}(a)(b) we show the evolution of the states on the Bloch sphere for the measurements $ M_0, M_1 $.  We see that $ M_0 $ has the effect of driving all states towards the south pole of the Bloch sphere, while $ M_1 $ drives all states to the north pole, following longitudinal lines.  This is consistent with the imaginary time operator $ e^{\pm \epsilon Z } $, as given in (\ref{measexp0}) and (\ref{measexp}).  In Fig. \ref{fig2}(c), we show the fidelity $ F = | \langle \psi_t | E_0 \rangle |^2  $ for three different measurement sequences.  Due to the randomness of quantum measurements, each sequence gives a different trajectory, but all cases converge to the ground state $ | E_0 \rangle = | 1 \rangle $.  Averaging over many random trajectories yields a smooth exponential curve approaching the target state.  A semilog plot (Fig. \ref{fig2}(c) inset) verifies the exponential evolution, consistent with ITE.  In fact, for this case it can be shown exactly that any trajectory is equivalent to applying a power of $ M_0 $ which approaches the ground state (see Supplementary Material).  In Fig. \ref{fig2}(d) we plot the peak position of the $ A$-function for the same three trajectories as in Fig. \ref{fig2}(c).  We see that there are broadly two regimes where there is a random movement of the peak position, followed by a region of stability, where the Gaussian approaches the ground state energy.  In the initial random evolution, when $ x^{\max}_{k_0 k_1} > \epsilon E_{\text{th}} $, several spin flips induced by $ U_C$ occur, until the random movement stabilizes to the correct energy range.  After the correct peak position is established, the fidelity quickly evolves towards the ground state.

\begin{figure}[t]
\includegraphics[width=\linewidth]{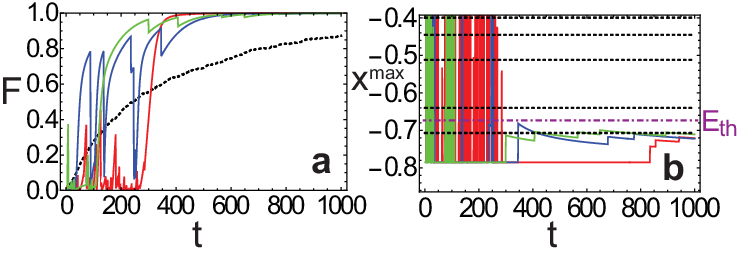}
\caption{(a) The fidelity of the state with respect to the ground state of the $ L = 5 $ site transverse Ising model with $ \lambda = 1 $ and $ \epsilon = 0.12 $ after $ t $ rounds of measurement and correction under  (\ref{generalevolution}) for three random initial states  (solid lines). Dashed lines show the averaged fidelity of 1000 evolutions.  (b) The peak position $ x_{k_0 k_1}^{\max} $ as defined in (\ref{gaussmax}) of the function $ A$ (solid lines).  Dashed lines show the energy eigenstates $ \epsilon E_n $ and the dashed dotted line $ E_{\text{th}} $. }
\label{fig3} 
\end{figure}

\paragraph{Example 2: Transverse-field Ising model}

We next show an example of the transverse-field Ising model with the Hamiltonian
$ H^{(1)}  = \lambda \sum_{n=1}^L X_n, H^{(2)} = \sum_{n=1}^{L-1} Z_n Z_{n+1}, H  = H^{(1)} + H^{(2)} $. Here, $ L$ is the number of qubits in the chain, and we take $ E_{\text{th}} = (E_0 + E_1)/2 $. We assume that each of the terms in the Hamiltonian must be implemented separately to construct the measurement operators.  We perform a second order Suzuki-Trotter expansion with $ M_n = \langle n |_a e^{-i \epsilon H^{(1)} \otimes Y /2 } e^{-i \epsilon H^{(2)} \otimes Y } e^{-i \epsilon H^{(1)} \otimes Y /2 } | + \rangle_a $ (see Supplementary Materials).  The conditional operator is chosen to be a random local unitary $ U_C = \otimes_{n=1}^L e^{2 \pi i (\phi_n^x X_n + \phi_n^y Y_n + \phi_n^z Z_n)} $, where  $ \phi_n^\alpha \in [0,1] $. We show the fidelity of the procedure with respect to the target state in Fig. \ref{fig3}(a).  Again we see two stages where there is a random evolution of the fidelity, followed by a smoother time evolution once the peak of Gaussian amplitude function is in the correct range.  For longer chains we observe a longer period of random evolution before the correct energy range is established, after which the system quickly converges to the ground state.

\paragraph{Complexity estimate}  We now briefly discuss the complexity of the proposed algorithm. First, the measurements $ M_n $ can be typically performed efficiently for a given Hamiltonian using a Suzuki-Trotter decomposition (see Supplementary Information).  Due to the flexibility of the choice of the operator $ U_C $, this can also be typically be implemented efficiently.  The complexity of the algorithm then results from the number of measurements that need to be made in total.  Based on the behavior observed in Figs. \ref{fig2} and \ref{fig3}, we model the initial part of the measurement sequence as a stochastic process, where the algorithm repeats until the criterion $ x^{\max}_{k_0 k_1} < \epsilon E_\text{th} $ is satisfied (see Supplementary Material).  The number of required measurements until this occurs can be estimated by evaluating the probability of obtaining a sequence with $ k $ consecutive  $M_0 $ outcomes together with the average failed sequence length.  Although it is not easy to obtain a simple expression for the general case complexity, for two particular cases, assuming an initial state with equal superposition, it is possible to estimate the typical number of measurements before convergence.  These are Hamiltonians with (I) a uniform density of states and (II) a completely degenerate spectrum of excited states (see Supplementary Information). For (I), we obtain a scaling as 
$ O( 1/(\epsilon \Delta)^2 ) $, where $ \Delta = E_1 - E_0 $ is the gap.  We note that there is an implicit dependence upon system dimension in this relation, due to the 
requirement that $ -\pi/4 \le \epsilon E_n \le \pi/4 $.  For example, for an exponential number of states, $ \epsilon \Delta $ is exponentially vanishing and the final scaling increases exponentially for unstructured problems. For (II), we find that the scaling is $ O(D) $, where $ D $ is the system dimension.

\paragraph{Conclusions} 

We have proposed a method of performing deterministic ITE, using measurements and conditional unitary operations. Due to use of quantum measurements, the evolution is stochastic within Hilbert space on a shot-to-shot basis.  Averaging over trajectories reveals an exponential evolution that is consistent with ITE. The approach is generic, one does not need to know the ground state  before executing the algorithm, and the measurement operators can be constructed 
with a Suzuki-Trotter decomposition so that it is compatible with gate based quantum computing. The measurement operators and unitary operators can be constructed efficiently,  but the number of measurements that need to be performed before convergence depends upon the nature of the Hamiltonian and the initial state. The algorithm is guaranteed to only converge if the energy of the state is lower than $ E_{\text{th}} $. 


The algorithm that we present here can be considered a generalization of several related works which use the same basic framework.  For example in Ref. \cite{ilo2023measurement} a similar method was proposed to generate supersinglet states, and also maximally entangled states of atomic ensembles in Ref. \cite{chaudhary2023macroscopic}.  A four-qubit linear graph state was also deterministically generated using the method in Ref. \cite{kondappan2023imaginary}. We have found that the algorithm converges to the ground state for every problem Hamiltonian that we have given it.   In our algorithm, we chose a relatively simple strategy for the adaptive unitary operator (\ref{udef}) where the state is rotated if the measurement outcomes do not fall in the targeted range.  Since $ E_{k_0 k_1}^{\max} $ is an energy estimate of the state, more complex strategies to rotate the state to the ground state could be made.   Another potential improvement is to choose a judicious initial state to improve the convergence of the scheme. 

This work is supported by the National Natural Science Foundation of China (62071301); NYU-ECNU Institute of Physics at NYU Shanghai; Shanghai Frontiers Science Center of Artificial Intelligence and Deep Learning; the Joint Physics Research Institute Challenge Grant; the Science and Technology Commission of Shanghai Municipality (19XD1423000,22ZR1444600); the NYU Shanghai Boost Fund; the China Foreign Experts Program (G2021013002L); the NYU Shanghai Major-Grants Seed Fund; Tamkeen under the NYU Abu Dhabi Research Institute grant CG008; and the SMEC Scientific Research Innovation Project (2023ZKZD55). J.S. is supported by the National Natural Science Foundation of China Grant Nos. 11925507 and 12047503. 


\section{Supplementary Information}

\setcounter{figure}{0}
\setcounter{equation}{0}
\makeatletter
\renewcommand{\thefigure}{S\@arabic\c@figure}
\renewcommand{\theequation}{S\@arabic\c@equation}

\newcommand{\manuallabel}[2]{\def\@currentlabel{#2}\label{#1}}
\makeatother
\manuallabel{fig1}{1}
\manuallabel{fig2}{2}
\manuallabel{fig3}{3}

\section{Construction of measurement operators}
\label{sec:constructionmeas}

\subsection{Derivation of Eqs. (2) and (3)}

To realize the measurement operator in Eqs. (2) and (3) of the main text, we use an ancilla qubit to perform
a weak measurement of the energy via a Hamiltonian of the form $ H \otimes Y $.  The ancilla qubit is initially prepared in the state $ |+\rangle_a = ( | 0 \rangle_a + | 1 \rangle_a)/\sqrt{2} $, where $ Y $ is the Pauli-$Y$ matrix. Interacting the total Hamiltonian for a time $ \epsilon $ we have
\begin{align}
e^{- i H \otimes Y \epsilon } &  | \psi_0 \rangle |+ \rangle_a  = \sum_n e^{-iE_n \epsilon Y} 
\langle E_n | \psi_0 \rangle   | E_n \rangle |+ \rangle \nonumber \\
& = \frac{1}{\sqrt{2}} \sum_n \langle E_n | \psi_0 \rangle  | E_n \rangle  \Big[ ( \cos E_n \epsilon - \sin E_n \epsilon) |0 \rangle_a \nonumber \\
& + ( \cos E_n \epsilon + \sin E_n \epsilon) |1 \rangle_a \Big].  
\end{align}
Projecting on the ancilla onto the $ |0 \rangle_a $ state gives
\begin{align}
& |0 \rangle_a \langle 0 |_a  e^{- i H \otimes Y \epsilon } | \psi_0 \rangle |+ \rangle_a \nonumber \\
& = \frac{1}{\sqrt{2}} \sum_n \langle E_n | \psi_0 \rangle  ( \cos E_n \epsilon - \sin E_n \epsilon) | E_n \rangle |0 \rangle_a
\end{align}
while the $ |1 \rangle_a $ outcome gives 
\begin{align}
& |1 \rangle_a \langle 1 |_a  e^{- i H \otimes Y \epsilon } | \psi_0 \rangle |+ \rangle_a \nonumber \\
& = \frac{1}{\sqrt{2}} \sum_n \langle E_n | \psi_0 \rangle  ( \cos E_n \epsilon + \sin E_n \epsilon) | E_n \rangle |1 \rangle_a  .  
\end{align}
The ancilla qubit decouples from the system after the measurement and the combined effect can be given according to the expressions given in Eq. (2) and (3) of the main text.

\subsection{Suzuki-Trotter decomposition}

The approach of the previous section can be used to construct the measurement operator if the Hamiltonian $ H \otimes Y $ is readily implementable.  For complex Hamiltonians involving many terms $ H = \sum_n H^{(n)} $, it may be necessary to construct the total Hamiltonian evolution via a Suzuki-Trotter decomposition.  We show in this section the measurement operators for this case.  

For a Hamiltonian consisting of two non-commuting terms we may perform a first order Suzuki-Trotter decomposition to give the measurement operator
\begin{align}
M_n & = \langle n |_a e^{-i H_0^{(1)} \otimes Y \epsilon } e^{-i H_0^{(2)} \otimes Y \epsilon }  |+ \rangle_a \nonumber \\
& = \frac{1}{\sqrt{2}} ( \cos \epsilon H^{(1)} \cos \epsilon H^{(2)} - \sin \epsilon H^{(1)} \sin \epsilon H^{(2)}) \nonumber \\
& - (-1)^n \frac{1}{\sqrt{2}} ( \cos \epsilon H^{(1)} \sin \epsilon H^{(2)} + \sin \epsilon H^{(1)} \cos \epsilon H^{(2)}) 
\label{firstorder}
\end{align}
for $ n \in \{0,1\} $. This has errors at the level of $ O(\epsilon^2) $.  For the transverse Ising model in Example 2 of the main text, we have found that the above first order Suzuki-Trotter decomposition did not have good convergence properties.  We instead used the second order Suzuki-Trotter decomposition
\begin{align}
M_n  = & \langle n |_a e^{-i H_0^{(1)} \otimes Y \epsilon/2 } e^{-i H_0^{(2)} \otimes Y \epsilon } e^{-i H_0^{(1)} \otimes Y \epsilon/2 } |+ \rangle_a \nonumber \\
= & \frac{1}{\sqrt{2}} ( \cos \frac{\epsilon H^{(1)}}{2} \cos \epsilon H^{(2)} \cos \frac{\epsilon H^{(1)}}{2}  \nonumber \\
& -\cos \frac{\epsilon H^{(1)}}{2} \sin \epsilon H^{(2)} \sin \frac{\epsilon H^{(1)}}{2} \nonumber \\
& -\sin \frac{\epsilon H^{(1)}}{2} \cos \epsilon H^{(2)} \sin \frac{\epsilon H^{(1)}}{2} \nonumber \\
& -\sin \frac{\epsilon H^{(1)}}{2} \sin \epsilon H^{(2)} \cos \frac{\epsilon H^{(1)}}{2} ) \nonumber \\
& - (-1)^n \frac{1}{\sqrt{2}} ( \cos \frac{\epsilon H^{(1)}}{2} \cos \epsilon H^{(2)} \sin \frac{\epsilon H^{(1)}}{2}  \nonumber \\
& + \cos \frac{\epsilon H^{(1)}}{2} \sin \epsilon H^{(2)} \cos \frac{\epsilon H^{(1)}}{2} \nonumber \\
& + \sin \frac{\epsilon H^{(1)}}{2} \cos \epsilon H^{(2)} \cos \frac{\epsilon H^{(1)}}{2} \nonumber \\
& -\sin \frac{\epsilon H^{(1)}}{2} \sin \epsilon H^{(2)} \sin \frac{\epsilon H^{(1)}}{2} )
\label{secondorder}
\end{align}
for $ n \in \{0,1\} $.  This has errors at the level of $ O(\epsilon^3) $. The superior convergence of the second order Suzuki-Trotter form is attributed to the fact that $ M_0 $ and $ M_1 $ have the same eigenstates for (\ref{secondorder}) but not (\ref{firstorder}), resulting in better stability of the fixed point.

\section{Ground state as the unique fixed point of evolution}
\label{sec:groundfixed}

Assume that $ E_0 < E_\text{th} < E_1 $, where $E_0 $ is the ground state energy and $ E_1 $ is an excited state that is not degenerate with the ground state, i.e. $ E_1 > E_0 $. First limiting ourselves to energy eigenstates $ |E_n \rangle $, we show that only the ground state $ |E_0 \rangle $ (and its degenerate states) are fixed points of the iteration.  We then later generalize to more general states and show that only energy eigenstates need to be considered.  

First consider the case of the initial state being the ground state $ | \psi_0 \rangle = | E_0 \rangle $.  After $ K = k_0 + k_1 $ applications of the measurement operator one obtains
\begin{align}
U_{k_0 k_1} M_0^{k_0} M_1^{k_1} |E_0 \rangle = A_{k_0 k_1}(\epsilon E_0) U_{k_0 k_1} |E_0 \rangle,
\label{groundstateappl}
\end{align}
where we used Eq. (4) in the main text.  To see the effect of the unitary $ U_{k_0 k_1} $, consider the most likely outcome of the measurement sequence 
$ k_0 \approx p_0 K, k_1 \approx p_1 K $ where $ K $ is the total number of measurements. Here, the probabilities of obtaining the two measurement outcomes for various eigenstates is
\begin{align} 
p_0 & = \langle E_n | M_0^\dagger M_0 | E_n \rangle = \frac{1}{2} ( 1 - \sin 2 \epsilon E_n) \nonumber \\
p_1 & = \langle E_n | M_1^\dagger M_1 | E_n \rangle = \frac{1}{2} ( 1 + \sin 2 \epsilon E_n)  . 
\end{align}
Substituting these values into Eq. (6) of the main text, this gives 
\begin{align}
x^{\max}_{k_0 k_1} & =  \frac{1}{2} \arcsin \left( \frac{k_1-k_0}{k_0+k_1} \right) \nonumber \\
& =  \frac{1}{2} \arcsin \left( p_1-p_0 \right)  \nonumber \\
& = \epsilon E_0 .
\label{groundstateconv}
\end{align}
We see that this obeys $ x^{\max}_{k_0 k_1} = \epsilon E_0  < \epsilon E_{\text{th}} $ such that $U_{k_0 k_1} = I  $ according to Eq. (8) of the main text. Therefore in this case
\begin{align}
U_{k_0 k_1} M_0^{k_0} M_1^{k_1} |E_0 \rangle \propto |E_0 \rangle,
\label{groundstateappl2}
\end{align} 
and the ground state is a fixed point of the evolution.  

For any initial state that is an excited state $ |E_n \rangle $ with $ n > 0 $, using similar arguments to (\ref{groundstateconv}), the measurement readouts converge to $ x^{\max}_{k_0 k_1} = \epsilon E_n $.  However, if  $ x^{\max}_{k_0 k_1} \ge \epsilon E_\text{th} $, the state is rotated away from $ | E_n \rangle $, since by definition $ |\langle E_n | U_C | E_m \rangle | > 0, \forall n,m $, and the measurement operators are diagonal in the energy basis.   Hence any excited state is not a fixed point of the iteration.    

Now consider the more general case of an arbitrary state.  Since the measurement operators are diagonal in the energy basis, for the case that $ x^{\max}_{k_0 k_1} < \epsilon E_\text{th} $ where $ U_{k_0 k_1}  = I $, the ground state is the only energy eigenstate which is a fixed point.  Then the only possibility is that the state $|\psi\rangle$ is an eigenstate of the combination of $ U_C $ and the measurement operators
\begin{align}
U_C M_0 | \psi \rangle &  \propto | \psi \rangle \nonumber \\
U_C M_1 | \psi \rangle  & \propto | \psi \rangle  ,
\end{align}
where we must consider both possibilities since in general  either outcome may occur. 
Such a state must produce the same state (up to a proportionality factor) for either measurement operator. For an arbitrary state we may evaluate 
\begin{align}
U_C M_0 | \psi \rangle & = \sum_n \langle E_n | \psi \rangle ( \cos \epsilon E_n - \sin \epsilon E_n ) | C_n \rangle \nonumber \\
U_C M_1 | \psi \rangle & = \sum_n \langle E_n | \psi \rangle ( \cos \epsilon E_n + \sin \epsilon E_n )  | C_n \rangle
\end{align}
where $ |C_n \rangle = U_C | E_n \rangle $ are a set of orthogonal basis states.  It is only possible to have $ U_C M_0 | \psi \rangle \propto U_C M_1 | \psi \rangle $ if $ | \psi \rangle $ is one of the energy eigenstates, which removes the $n$ dependence and hence the factor of $ \cos \epsilon E_n \pm  \sin \epsilon E_n $ after normalization. Since we have already shown that the only energy eigenstate that is a fixed point of the evolution is $ |E_0 \rangle $, this completes the proof.

\section{Complexity estimate of the algorithm}

We now estimate the complexity of the algorithm for  two prototypical Hamiltonian spectra.  Firstly, the measurement operators $ M_n $ can be constructed efficiently using the methods given in the first section of supplementary information.  Even in the case that a Suzuki-Trotter decomposition is used to construct the measurement operators, this will scale polynomially with the number of terms in the Hamiltonian, which  typically scales polynomially with the number of qubits.  In this way the measurement operators can be constructed efficiently.  The $U_C $ operator can also be chosen according to what gates are available such that it satisfies $ |\langle E_n | U_C | E_m \rangle | > 0, \forall n,m $.  Thus the unitary correction operator can also be chosen to be implemented efficiently with the number of qubits.  

The main complexity of the algorithm is then determined by the number of iterations is required before convergence is attained.  A hint of the dynamics towards convergence can be seen in Figs.  2 and 3 of the main text.  We see here that initially there is a period of chaotic evolution of the fidelity and the peak position.  At some number of iterations, the measurement sequence ``locks in'', after which rapid convergence towards a fidelity of 1 is attained.  

We can understand the dynamics in the following way.  We wish to obtain imaginary time evolution by taking advantage of the similarity of $ M_0 $ with the operator $ e^{-H \epsilon} $.  However, since measurements are random, we are not guaranteed to obtain only $ M_0 $ and one will obtain $ M_1 $ with some probability.  The algorithm as given in Eqns. (7)-(9) of the main text keeps trying until one obtains a sequence satisfying $ x_{k_0 k_1}^{\max} < \epsilon E_{\text{th}} $, which amounts to convergence of a low energy state below a particular threshold.  The algorithm keeps iterating until one obtains the desired sequence where there is a sufficiently large number of $ M_0 $ measurements over $ M_1 $ measurements. 

For a state that has a uniform amplitude in energy eigenstates (i.e. $ |\langle \psi | E_n \rangle| = $ constant), the probability of obtaining $ M_0 $ and $ M_1 $ is $\approx 1/2$.  This might naively suggest that to obtain $ k $ measurements that are all $ M_0 $ has a probability of $ 1/2^k $, which is vanishing for a large number of measurements.  This is however a large underestimate, due to the way quantum mechanical measurements work.  In fact, getting the $ M_0 $ outcome is a self-enhancing process, where each time an $ M_0 $ outcome occurs, it becomes more likely to obtain $ M_0 $ again.

	\subsection{Uniform density of states}
	\label{sec:uniform}

To see this quantitatively, let us consider the following example. We choose a problem in a $ D $ dimensional Hilbert space with uniformly distributed energies.  The Hamiltonian is
	\begin{align}
	H = \sum_{n=0}^{D-1} E_n  | n \rangle \langle n | 
	\label{uniham}
	\end{align}
	where for a uniform distribution we take
	\begin{align}
	E_n =  2n - D+1
	\label{uniham2}
	\end{align}
	such that the ground state $ E_0 = -(D-1) $ and the highest energy state is $ E_{D-1} = D-1 $.   We may choose $ \epsilon $ such that the spectrum occupies the full range of the domain of $ A_{k_0 k_1} (x) $, which is $ - \pi/4 \le x \le \pi/4 $.   Namely, $ \epsilon = \pi/(4(D-1)) $  such that $ \epsilon E_0 = -\pi/4 $ and $ \epsilon E_{D-1} = \pi/4 $.   This maximizes $ \epsilon $ and typically gives the best performance of the algorithm. As discussed in the main text, we require choosing the energy threshold $ E_{\text{th}} < (E_0 + E_1)/2 $ such that the ground state has the largest amplification factor. With this choice, under the Gaussian approximation of the function $ A_{k_0 k_1} (x) $, the amplitude function obeys $  A_{k_0 k_1} (\epsilon E_0 ) > A_{k_0 k_1} (\epsilon E_n ) $ for $ n > 0 $ and the ground state is amplified according to Eq. (4) of the main text.   We consider a case with $ D \gg 1 $ and a uniform density of states, such that we may take $ E_{\text{th}} \approx E_0 $, which means that the condition in Eq. (8) of the main text is $ x_{k_0 k_1}^{\max} < -\pi/4 $. Translating this to $ k_0, k_1 $ using Eq. (6) of the main text, this means that one must obtain {\it every} measurement outcome to be $ M_0 $, and hence $ k_1 = 0 $ to satisfy the convergence. 

Let us calculate the probability of obtaining $ k $ measurement outcomes that are $M_0 $. For simplicity, suppose the initial state is in the state
\begin{align}
| \psi_0 \rangle = \frac{1}{\sqrt{D}} \sum_{n=0}^{D-1} | n \rangle . 
\label{inistate}
\end{align}
Then after $ k $ outcomes for the measurement $ M_0 $, the resulting state is 
\begin{align}
M_0^k | \psi_0 \rangle = \frac{1}{\sqrt{D}} \sum_{n=0}^{D-1} \cos^k ( \epsilon E_n  + \pi/4)  | n \rangle .  
\label{mzerok}
\end{align}
The probability of this measurement outcome, where there are $ k $ consecutive $ M_0 $ outcomes, is 
\begin{align}
p_k & = \langle \psi_0 | (M_0^\dagger)^k M_0^k | \psi_0 \rangle \nonumber \\
& = \frac{1}{D} \sum_{n=0}^{D-1} \cos^{2k} ( \epsilon E_n  + \pi/4) .  
\label{pkexact}
\end{align}
Let us assume a uniform density of states in energy space, such that the sum can be approximated by 
\begin{align}
p_k &  \approx \frac{1}{\pi/2} \int_{-\pi/4}^{\pi/4} dx \cos^{2k} ( x  + \pi/4) \nonumber \\
& = \frac{1}{4^k} \binom{2k}{k} \label{zerozeroprobexact}  \\
& \approx \frac{1}{\sqrt{k \pi}} ,
\label{zerozeroprob}
\end{align}
where the last approximation is valid for $ k \gg 1 $.  We observe that the probability of obtaining consecutive $ M_0 $ outcomes is in fact much larger than the $ 1/2^k $ estimate one would naively make from 
an independent probability assumption. 


The self-enhancing effect of consecutive $ M_0 $ outcomes is a weak measurement version of the familiar effect known for projective measurements.  For example, consider the projective measurements $ P_0 = | 0 \rangle \langle 0 | $ and $ P_1 = I - | 0\rangle \langle 0  | $.  If the measurement outcome $ P_0 $ occurs on an initial state $ P_0 | \psi_0 \rangle \propto | 0 \rangle $, then with unit probability subsequent outcomes will all be $ P_0 $. In the weak measurement case that we consider, the enhancement of probability is more gradually attained, rather than after a single measurement. 

Returning to the case that we consider here, any time the measurement outcome $ M_1 $ is obtained, we violate the criterion  $ x_{k_0 k_1}^{\max} < \epsilon E_{\text{th}} = \epsilon E_0 = -\pi/4 $, and a new attempt at convergence begins.  We may estimate the total number of measurements before a sequence $M_0^k $ is obtained as $ 1/p_k $, multiplied by the average length of a failed sequence.  A failed sequence consists of any sequence which is shorter than the target length and ends in an 
$ M_1 $ outcome.  Specifically, if we consider $ M_0^k $  to be a successful sequence, then the sequences $ M_1, M_1 M_0, M_1 M_0^2, \dots, M_1 M_0^{k-1} $ to be the associated failed sequences.  To evaluate the probability of a failed measurement sequence, first evaluate the probability of obtaining $ M_0 $ following $ M_0^k$ as
\begin{align}
p_{0|k} = \frac{p_{k+1}}{p_k} = 1 - \frac{1}{2(k+1)} ,
\end{align}
where we used (\ref{zerozeroprobexact}).  The probability of obtaining $ M_1 $ following $ M_0^k$ is then
\begin{align}
p_{1|k} = 1- p_{0|k} = \frac{1}{2(k+1)} . 
\end{align}
The probability of obtaining an outcome $ M_1 M_0^{k-1} $ is then 
\begin{align}
p_{1|k-1} p_{k-1} =  \frac{2}{k 4^k} \binom{2k-2}{k-1} .
\end{align}
The total failure probability for a sequence of length $ k $ is then 
\begin{align}
p_k^{\text{fail}} = \sum_{k'=1}^k p_{1|k'-1} p_{k'-1} \nonumber \\
= 1 - \frac{1}{4^k} \binom{2k}{k}
\end{align}
which is equal to $ 1- p_k $ as expected.  The average length of a failed sequence is then
\begin{align}
T_k^{\text{fail}} & = \sum_{k'=1}^k k' p_{1|k'-1} p_{k'-1} \nonumber \\
& = \frac{k}{4^k} \binom{2k}{k} \nonumber \\
& \approx \sqrt{\frac{k}{\pi}} 
\end{align}
The expected total number of measurements before a target sequence of $ M_0^k $ is obtained is then 
\begin{align}
T= \frac{T_k^{\text{fail}}}{p_k } + k \approx 2k , 
\label{numberofmeas}
\end{align}
where we have added a $ k $ to the total to account for the number of measurements in $ M_0^k $ itself.

Now let us estimate what $ k $ should be such that the ground state is obtained with high fidelity. For large $ k $, we may approximate (\ref{mzerok}) with a Gaussian such that 
\begin{align}
M_0^k | \psi_0 \rangle = \frac{1}{\sqrt{D}} \sum_{n=0}^{D-1} \exp (-\frac{ (\epsilon E_n  + \pi/4)^2}{2 \sigma^2} )  |n \rangle .  
\end{align}
where the standard deviation is $ \sigma = 1/\sqrt{k} $. To obtain a high fidelity convergence to the ground state, we require that at the energy of the first excited state $ E_1 $, the Gaussian sufficiently suppresses its amplitude.  Hence we require
\begin{align}
\epsilon (E_1 - E_0) \sim 2 \sigma = \frac{2}{\sqrt{k}} . 
\label{highfidcrit}
\end{align}
where we have put a $ 2 \sigma $ standard deviation which suppresses the first excited state by a factor 0.14.  

Putting together (\ref{numberofmeas}) and (\ref{highfidcrit}) we obtain the expected total number of measurements
\begin{align}
T \approx \frac{8}{(\epsilon \Delta)^2} ,
\label{finalscaling}
\end{align}
where $ \Delta = E_1-E_0 $.

	Naively, the uniform density of states Hamiltonian is a constant gap problem with $ \Delta = 2 $, and is independent of $ D $, according to Eq. (\ref{uniham2}).  Hence it appears that the scaling of the imaginary time evolution allows for a way to solve the problem independent of problem dimension $ D $, which we consider to be an exponentially large quantity.   However, an important point is that we must also  choose an $ \epsilon $ such that the energy spectrum of the Hamiltonian lies in the range $ -\pi/4 \le \epsilon E_n \le \pi/4 $.  As obtained previously, here we require $ \epsilon \le \pi/(4(D-1)) $.   This means that even if the gap $ \Delta $ itself is a constant as the problem size $ D $  grows, we must fit an exponential number of states within the range $ -\pi/4 \le \epsilon E_n \le \pi/4 $. Therefore the combination $ \epsilon \Delta $ is an exponentially small quantity for exponentially large $ D $.  Then according to our scaling $ O(1/(\epsilon \Delta)^2 )$ as written in the main text, the algorithm would take an exponential time for convergence.

\begin{figure}[t!]
		\centering
		\includegraphics[width=\linewidth]{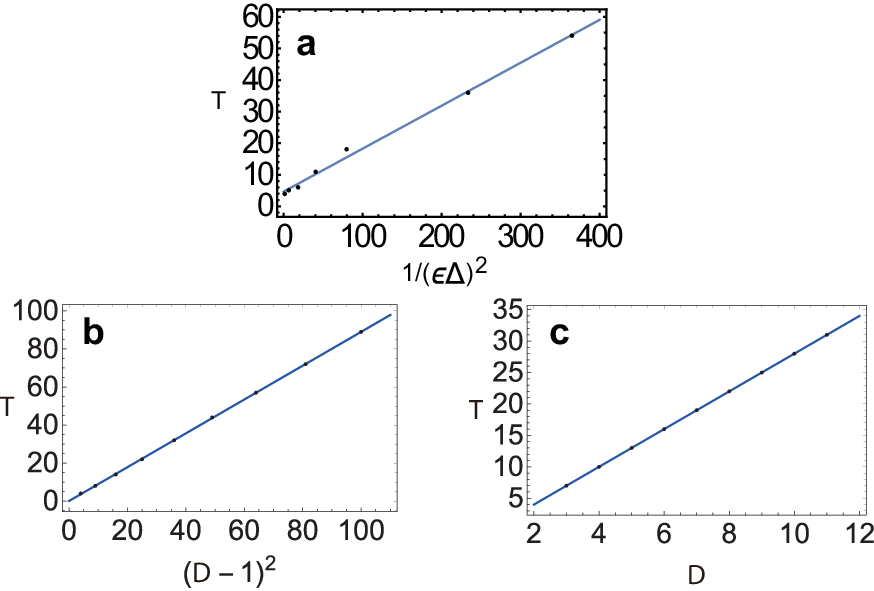}
		\caption{ Numerical evaluations of the time scaling of the measurement-based imaginary time algorithm.  Points shows the most probable number of measurements to obtain a fidelity of 0.9, and solid lines shows a linear best fit.  (a) The time scaling for the uniform density of states Hamiltonian (\ref{uniham}) with fixed $ D = 3 $ and varying $ \epsilon $.  The initial state is (\ref{inistate}). 
				(b) As with (a), but for $ \epsilon = \pi/(4(D-1)) $ and varying $ D $. (c) The scaling of golf course energy landscape (\ref{golf}). We take  $\epsilon=\frac{\pi}{4}$ and varying  $ D$.   }
		\label{figscalingandpk}
	\end{figure}

	We have performed some numerical analysis to verify the above scaling.  In Fig. \ref{figscalingandpk} we show the most probable number of measurements to first attain a fidelity of 0.9.  To obtain this, we run the algorithm as described in the main text applied to the Hamiltonian (\ref{uniham}) many times and find what is the most likely number of measurements.    In the case of Fig. \ref{figscalingandpk}(a), we fix $ D $ and vary $ \epsilon $ in the range 
	$0 <  \epsilon \le \frac{\pi}{4(D-1)} $.  We see that the numerical values follow the predicted scaling $ T \propto 1/(\epsilon \Delta)^2 $.  For Fig. \ref{figscalingandpk}(b), we take $ \epsilon = \frac{\pi}{4(D-1)} $ and vary $ D $.  In terms of $ D $, we therefore expect $ T \propto (D-1)^2 $ from (\ref{finalscaling}).  We see excellent agreement to the numerical data, again taking the most probable number of measurements to obtain a fidelity of 0.9.  
	
	The Hamiltonian (\ref{uniham}) does not appear to be a computationally difficult problem due to its rather simple energy structure.  It can however be converted to a computationally difficult problem by reassigning the energies randomly to the states.  In this case, the problem has no structure and would be a computationally hard problem.  Specifically, the Hamiltonian reads
	\begin{align}
	H = \sum_{n=0}^{D-1} E_n  | P(n) \rangle \langle P(n) | 
	\label{combham}
	\end{align}
	where $ P(n) $ is a permuting function which rearranges the states in a randomized manner with no structure.  The energy distribution is uniform as before, given in Eq. (\ref{uniham2}).  
	In this case formally the analysis is the same as above except that the energy labels are reordered.  The analysis can be repeated in the same way, such that we obtain the same scaling. We obtain the same result since the simple energy structure was never exploited (in constructing $ U_C $ for example), hence the scaling remains the same.


We emphasize that the assumptions of the uniformly distributed energies and even superposition initial state are not required for the running of our algorithm itself. The assumptions are made for the purpose of the complexity analysis to obtain a simple expression.  
	Eqs. (4) and (5) in the main text show when a weak measurement sequence is applied on an arbitrary initial state, amplitude function takes a Gaussian form. As Fig. 1(a) in the main text shows, when $k_1=0, k_0=50$, $A_{k_0,k_1}(\epsilon E_n)$ monotonically decreases within the domain $\epsilon E_n \in [-\frac{\pi}{4},\frac{\pi}{4}]$. This means that the energy corresponding to $\epsilon E = -\frac{\pi}{4}$ will be the state which has the highest probability after many measurements. This is true as long as the ground state amplitude is not precisely zero, and is also independent of the energy structure of the problem.

	\subsection{Golf course energy landscape}
	
	Next we examine the opposite limit of a golf course Hamiltonian defined as
	\begin{align}
	H = - | 0 \rangle \langle 0 | + \sum_{n=1}^{D-1} | n \rangle \langle n | . 
	\label{golf}
	\end{align}
	Here the ground state is $ | 0 \rangle $ and has energy $ E_0 = -1 $, and the remaining states have an energy $ E_n = 1 $ for $ n \ge 1 $.  
	
	We again choose a full range of the domain of $ A_{k_0 k_1} (x) $ such that $ -\pi/4 \le x \le \pi/4 $, which means we choose $ \epsilon E_0 = -\pi/4 $ and $ \epsilon E_n = \pi/4 $ for $ n \ge 1 $, which means that $ \epsilon = \pi/4 $. In this case the measurement operators are according to (2) and (3) in the main text
	\begin{align}
	M_0 & = | 0 \rangle \langle 0 | \nonumber \\
	M_1 & = \sum_{n=1}^{D-1} |n \rangle \langle n | .  
	\label{measopsgolf}
	\end{align}
	
	Consider again a uniformly distributed initial state 
	\begin{align}
	|\psi_0 \rangle = \frac{1}{\sqrt{D}} \sum_{n = 0 }^{D-1} | n \rangle .  
	\end{align}
	The probability of obtaining the ground state in this case is
	\begin{align}
	p_0 = \langle \psi_0 | M_0^\dagger M_0 | \psi_0 \rangle  = \frac{1}{D} . 
	\end{align}
	
	The measurement operators (\ref{measopsgolf}) are in this case projective operators and are orthogonal $ M_0 M_1 = 0 $.  As such, after the first measurement outcome of either $ M_0 $ or $ M_1 $, all subsequent measurements are obtained with the same outcome.  Therefore, the length of a failed sequence is length $ T^{\text{fail}} = 1$, corresponding to the $ M_1 $ outcome alone.  
	
	Assuming that the corrective unitary $ U_C $ produces a state with amplitude $ \sim 1 / \sqrt{D} $ on the ground state after each failed sequence, the total number of measurements before obtaining an outcome $ M_0 $ is 
	\begin{align}
	T \approx \frac{T^{\text{fail}}}{p_0} + 1 = D+1 ,  
	\label{golfscaling}
	\end{align}
	where the additional 1 is the length of the successful sequence $ M_0 $.  
	
	We note that in a similar way to (\ref{combham}), the golf course Hamiltonian can be a computationally hard problem by permuting the state labels such that the lowest energy state is not necessarily the state $ | 0 \rangle $.

The above scaling was verified numerically by directly running the algorithm using the Hamiltonian (\ref{golf}), as shown in Fig. \ref{figscalingandpk}(c).  We again see excellent agreement with the theoretical prediction of (\ref{golfscaling}). Considering $ D$ to be an exponentially large quantity, the time scaling of the problem is therefore exponential.

	\subsection{Other problems}
	
	Here we make a brief comment regarding more complex Hamiltonian problems that occur in combinatorial optimization problems such as 3SAT or MAXCUT. Such problems have a gap that is constant with respect to the system dimension, in a similar way to (\ref{combham}).  Such problems can be considered to be an intermediate case between the uniform density of states and the golf course energy landscape, since there may exist a high level of degeneracy particularly in the middle of the spectrum.  We have seen that in both limiting cases, there is an exponential overhead.  Hence we expect that in such combinatorial problems the time scaling remains exponential.

\section{Convergence error of the algorithm}

We now estimate the error attained by the algorithm during convergence of the algorithm.  For a sufficiently large number of measurements, the algorithm converges to a $ k_0, k_1 $ such that $ x_{k_0 k_1}^{\max} < \epsilon E_{\text{th}} $, so that the unitary $ U_{k_0 k_1} = I $ as described in the previous section.  The more the function $ A $ is peaked at an energy less than $ x_{k_0 k_1}^{\max} = \epsilon E_{\text{th}} $, the better the fidelity is, since the Gaussian form tends to suppress high energy states.  Hence the worst-case fidelity is when $ x_{k_0 k_1}^{\max} = \epsilon E_{\text{th}} $, so that it barely satisfies the convergence threshold.  

Now consider for simplicity a non-degenerate ground and first excited state, separated by an energy gap $ \Delta$.  Let us also parametrize $ E_{\text{th}} = E_0 + \delta  $.  According to Eq. (4) in the main text, the resulting unnormalized state is 
\begin{align}
M_0^{k_0} M_1^{k_1} | \psi_0 \rangle = \sum_n A_{k_0 k_1} ( \epsilon E_n) \langle E_n | \psi_0 \rangle | E_n \rangle
\label{measurestate}
\end{align}
where we can approximate the $A$-function by a Gaussian of form
\begin{align}
A_{k_0 k_1} ( x) \propto e^{- K  (x-x_{k_0 k_1}^{\max})^2  }, 
\end{align}
where $ K = k_0 + k_1 $ is the total number of measurements.  The amplitude factors on the ground and first excited states are
\begin{align}
A_{k_0 k_1} ( \epsilon E_0 ) & \propto e^{- K \epsilon^2  \delta^2} \nonumber \\
A_{k_0 k_1} ( \epsilon E_1 ) & \propto e^{- K \epsilon^2  ( \Delta - \delta )^2}  . 
\end{align}
The fidelity with the ground state for the unnormalized state (\ref{measurestate}) is 
\begin{align}
F & = \frac{|A_{k_0 k_1} ( \epsilon E_n) \langle E_0 | \psi_0 \rangle |^2}{\sum_{n=0}^D  | A_{k_0 k_1} ( \epsilon E_n) \langle E_n | \psi_0 \rangle|^2 } \nonumber \\
& = \frac{1}{1+ \sum_{n=1}^D  \left| \frac{A_{k_0 k_1} ( \epsilon E_n) \langle E_n | \psi_0 \rangle}{A_{k_0 k_1} ( \epsilon E_0) \langle E_0 | \psi_0 \rangle} \right|^2 } ,
\end{align}
where $ D $ is the Hilbert space dimension. Due to the fact that $ A $ function is a Gaussian, assuming the initial coefficients $ \langle E_n | \psi_0 \rangle $ are of the same order, the fidelity can be estimated as
\begin{align}
F & \approx \frac{1}{1+  \left| \frac{A_{k_0 k_1} ( \epsilon E_1) \langle E_1 | \psi_0 \rangle}{A_{k_0 k_1} ( \epsilon E_0) \langle E_0 | \psi_0 \rangle} \right|^2 } \nonumber \\
& \approx 1- \left| \frac{A_{k_0 k_1} ( \epsilon E_1) \langle E_1 | \psi_0 \rangle}{A_{k_0 k_1} ( \epsilon E_0) \langle E_0 | \psi_0 \rangle} \right|^2 ,
\end{align}
assuming that $ \left| \frac{A_{k_0 k_1} ( \epsilon E_1) \langle E_1 | \psi_0 \rangle}{A_{k_0 k_1} ( \epsilon E_0) \langle E_0 | \psi_0 \rangle} \right|^2 \ll 1 $.  The error, or infidelity, is then 
\begin{align}
{\cal E} &  = 1 - F \nonumber \\
& \approx \left| \frac{\langle E_1 | \psi_0 \rangle }{\langle E_0 | \psi_0 \rangle} \right|^2
e^{- K \epsilon^2 \Delta ( \Delta   - 2 \delta )}
\end{align}
In order to converge to the ground state, we must set $ E_{\text{th}} < (E_0 + E_1)/2 $, which using our variables corresponds to $ \delta  < \Delta/2 $.  When this is an equality, the exponential is equal to 1, and there is an equal suppression factor of both the ground and first excited states.  Higher energy states are suppressed further.

\section{Exact evaluation of qubit imaginary time evolution}

The qubit example (Example 1 in the main text) allows for another way of understanding the dynamics.  In this case $ M_0 M_1 \propto I $, so that a general sequence involving multiple applications of $ U_C = X $ can be simplified as
\begin{align}
\prod_{t=1}^T \left( U_{k_0^{(t)} k_1^{(t)} }  M_{n_t} \right) |\psi_0 \rangle \propto  M_0^k X^{N_C} | \psi_0 \rangle 
\end{align}
where $ k $ is a non-negative integer and $ N_C $ is the number of the times $ U_C=X $ is applied. 
It can be ensured that $ k \ge 0 $ because whenever $ k_1 > k_0 $ (according to the criterion $ x_{k_0 k_1}^{\max} > \epsilon E_{\text{th}} = 0 $), $ X $ is applied, $ k_0 \leftrightarrow k_1 $ are interchanged, since $ X M_0 = M_1 X $, $ X M_1 = M_0 X $.  For example, we may simplify the sequence using these identities as
\begin{align}
X & M_1 M_1 X M_1 M_1 M_1 M_0 M_0 |\psi_0 \rangle \nonumber \\
& = M_0 M_0 M_1 M_1 M_1 M_0 M_0 X^2 |\psi_0 \rangle \nonumber \\
& \propto  M_0 |\psi_0 \rangle .  
\end{align}
In this way, it is possible to always ensure that $ k> 0 $, which converges towards the ground state for $k \gg 1 $.  


\begin{thebibliography}{72}
	\expandafter\ifx\csname natexlab\endcsname\relax\def\natexlab#1{#1}\fi
	\expandafter\ifx\csname bibnamefont\endcsname\relax
	\def\bibnamefont#1{#1}\fi
	\expandafter\ifx\csname bibfnamefont\endcsname\relax
	\def\bibfnamefont#1{#1}\fi
	\expandafter\ifx\csname citenamefont\endcsname\relax
	\def\citenamefont#1{#1}\fi
	\expandafter\ifx\csname url\endcsname\relax
	\def\url#1{\texttt{#1}}\fi
	\expandafter\ifx\csname urlprefix\endcsname\relax\def\urlprefix{URL }\fi
	\providecommand{\bibinfo}[2]{#2}
	\providecommand{\eprint}[2][]{\url{#2}}
	
	\bibitem[{\citenamefont{Sakurai and Commins}(1995)}]{sakurai1995modern}
	\bibinfo{author}{\bibfnamefont{J.~J.} \bibnamefont{Sakurai}} \bibnamefont{and}
	\bibinfo{author}{\bibfnamefont{E.~D.} \bibnamefont{Commins}},
	\emph{\bibinfo{title}{Modern quantum mechanics, revised edition}}
	(\bibinfo{year}{1995}).
	
	\bibitem[{\citenamefont{Magnus}(1954)}]{magnus1954exponential}
	\bibinfo{author}{\bibfnamefont{W.}~\bibnamefont{Magnus}},
	\bibinfo{journal}{Communications on pure and applied mathematics}
	\textbf{\bibinfo{volume}{7}}, \bibinfo{pages}{649} (\bibinfo{year}{1954}).
	
	\bibitem[{\citenamefont{Vidal}(2007)}]{vidal2007classical}
	\bibinfo{author}{\bibfnamefont{G.}~\bibnamefont{Vidal}},
	\bibinfo{journal}{Physical review letters} \textbf{\bibinfo{volume}{98}},
	\bibinfo{pages}{070201} (\bibinfo{year}{2007}).
	
	\bibitem[{\citenamefont{Lin et~al.}(2021)\citenamefont{Lin, Dilip, Green,
			Smith, and Pollmann}}]{lin2021real}
	\bibinfo{author}{\bibfnamefont{S.-H.} \bibnamefont{Lin}},
	\bibinfo{author}{\bibfnamefont{R.}~\bibnamefont{Dilip}},
	\bibinfo{author}{\bibfnamefont{A.~G.} \bibnamefont{Green}},
	\bibinfo{author}{\bibfnamefont{A.}~\bibnamefont{Smith}}, \bibnamefont{and}
	\bibinfo{author}{\bibfnamefont{F.}~\bibnamefont{Pollmann}},
	\bibinfo{journal}{PRX Quantum} \textbf{\bibinfo{volume}{2}},
	\bibinfo{pages}{010342} (\bibinfo{year}{2021}).
	
	\bibitem[{\citenamefont{Schuch et~al.}(2007)\citenamefont{Schuch, Wolf,
			Verstraete, and Cirac}}]{schuch2007computational}
	\bibinfo{author}{\bibfnamefont{N.}~\bibnamefont{Schuch}},
	\bibinfo{author}{\bibfnamefont{M.~M.} \bibnamefont{Wolf}},
	\bibinfo{author}{\bibfnamefont{F.}~\bibnamefont{Verstraete}},
	\bibnamefont{and} \bibinfo{author}{\bibfnamefont{J.~I.} \bibnamefont{Cirac}},
	\bibinfo{journal}{Physical review letters} \textbf{\bibinfo{volume}{98}},
	\bibinfo{pages}{140506} (\bibinfo{year}{2007}).
	
	\bibitem[{\citenamefont{McArdle et~al.}(2019)\citenamefont{McArdle, Jones,
			Endo, Li, Benjamin, and Yuan}}]{mcardle2019variational}
	\bibinfo{author}{\bibfnamefont{S.}~\bibnamefont{McArdle}},
	\bibinfo{author}{\bibfnamefont{T.}~\bibnamefont{Jones}},
	\bibinfo{author}{\bibfnamefont{S.}~\bibnamefont{Endo}},
	\bibinfo{author}{\bibfnamefont{Y.}~\bibnamefont{Li}},
	\bibinfo{author}{\bibfnamefont{S.~C.} \bibnamefont{Benjamin}},
	\bibnamefont{and} \bibinfo{author}{\bibfnamefont{X.}~\bibnamefont{Yuan}},
	\bibinfo{journal}{npj Quantum Information} \textbf{\bibinfo{volume}{5}},
	\bibinfo{pages}{1} (\bibinfo{year}{2019}).
	
	\bibitem[{\citenamefont{Chiofalo et~al.}(2000)\citenamefont{Chiofalo, Succi,
			and Tosi}}]{chiofalo2000ground}
	\bibinfo{author}{\bibfnamefont{M.~L.} \bibnamefont{Chiofalo}},
	\bibinfo{author}{\bibfnamefont{S.}~\bibnamefont{Succi}}, \bibnamefont{and}
	\bibinfo{author}{\bibfnamefont{M.}~\bibnamefont{Tosi}},
	\bibinfo{journal}{Physical Review E} \textbf{\bibinfo{volume}{62}},
	\bibinfo{pages}{7438} (\bibinfo{year}{2000}).
	
	\bibitem[{\citenamefont{Palpacelli et~al.}(2007)\citenamefont{Palpacelli,
			Succi, and Spigler}}]{PhysRevE.76.036712}
	\bibinfo{author}{\bibfnamefont{S.}~\bibnamefont{Palpacelli}},
	\bibinfo{author}{\bibfnamefont{S.}~\bibnamefont{Succi}}, \bibnamefont{and}
	\bibinfo{author}{\bibfnamefont{R.}~\bibnamefont{Spigler}},
	\bibinfo{journal}{Phys. Rev. E} \textbf{\bibinfo{volume}{76}},
	\bibinfo{pages}{036712} (\bibinfo{year}{2007}).
	
	\bibitem[{\citenamefont{Liu and Makri}(2005)}]{liu2005bohm}
	\bibinfo{author}{\bibfnamefont{J.}~\bibnamefont{Liu}} \bibnamefont{and}
	\bibinfo{author}{\bibfnamefont{N.}~\bibnamefont{Makri}},
	\bibinfo{journal}{Molecular Physics} \textbf{\bibinfo{volume}{103}},
	\bibinfo{pages}{1083} (\bibinfo{year}{2005}).
	
	\bibitem[{\citenamefont{Wick}(1954)}]{wick1954properties}
	\bibinfo{author}{\bibfnamefont{G.-C.} \bibnamefont{Wick}},
	\bibinfo{journal}{Physical Review} \textbf{\bibinfo{volume}{96}},
	\bibinfo{pages}{1124} (\bibinfo{year}{1954}).
	
	\bibitem[{\citenamefont{Peskin}(2018)}]{peskin2018introduction}
	\bibinfo{author}{\bibfnamefont{M.~E.} \bibnamefont{Peskin}},
	\emph{\bibinfo{title}{An introduction to quantum field theory}}
	(\bibinfo{publisher}{CRC press}, \bibinfo{year}{2018}).
	
	\bibitem[{\citenamefont{Majid}(1994)}]{majid1994q}
	\bibinfo{author}{\bibfnamefont{S.}~\bibnamefont{Majid}},
	\bibinfo{journal}{Journal of Mathematical Physics}
	\textbf{\bibinfo{volume}{35}}, \bibinfo{pages}{5025} (\bibinfo{year}{1994}).
	
	\bibitem[{\citenamefont{Vesely}(1994)}]{vesely1994computational}
	\bibinfo{author}{\bibfnamefont{F.~J.} \bibnamefont{Vesely}},
	\emph{\bibinfo{title}{Computational Physics}} (\bibinfo{publisher}{Springer},
	\bibinfo{year}{1994}).
	
	\bibitem[{\citenamefont{Lester~Jr and Hammond}(1990)}]{lester1990quantum}
	\bibinfo{author}{\bibfnamefont{W.~A.} \bibnamefont{Lester~Jr}}
	\bibnamefont{and} \bibinfo{author}{\bibfnamefont{B.~L.}
		\bibnamefont{Hammond}}, \bibinfo{journal}{Annual Review of Physical
		Chemistry} \textbf{\bibinfo{volume}{41}}, \bibinfo{pages}{283}
	(\bibinfo{year}{1990}).
	
	\bibitem[{\citenamefont{Jarrell}(1992)}]{jarrell1992hubbard}
	\bibinfo{author}{\bibfnamefont{M.}~\bibnamefont{Jarrell}},
	\bibinfo{journal}{Physical review letters} \textbf{\bibinfo{volume}{69}},
	\bibinfo{pages}{168} (\bibinfo{year}{1992}).
	
	\bibitem[{\citenamefont{Baroni and Moroni}(1999)}]{baroni1999reptation}
	\bibinfo{author}{\bibfnamefont{S.}~\bibnamefont{Baroni}} \bibnamefont{and}
	\bibinfo{author}{\bibfnamefont{S.}~\bibnamefont{Moroni}},
	\bibinfo{journal}{Physical review letters} \textbf{\bibinfo{volume}{82}},
	\bibinfo{pages}{4745} (\bibinfo{year}{1999}).
	
	\bibitem[{\citenamefont{Byrnes et~al.}(2004)\citenamefont{Byrnes, Loan, Hamer,
			Bonnet, Leinweber, Williams, and Zanotti}}]{byrnes2004hamiltonian}
	\bibinfo{author}{\bibfnamefont{T.}~\bibnamefont{Byrnes}},
	\bibinfo{author}{\bibfnamefont{M.}~\bibnamefont{Loan}},
	\bibinfo{author}{\bibfnamefont{C.}~\bibnamefont{Hamer}},
	\bibinfo{author}{\bibfnamefont{F.~D.} \bibnamefont{Bonnet}},
	\bibinfo{author}{\bibfnamefont{D.~B.} \bibnamefont{Leinweber}},
	\bibinfo{author}{\bibfnamefont{A.~G.} \bibnamefont{Williams}},
	\bibnamefont{and} \bibinfo{author}{\bibfnamefont{J.~M.}
		\bibnamefont{Zanotti}}, \bibinfo{journal}{Physical Review D}
	\textbf{\bibinfo{volume}{69}}, \bibinfo{pages}{074509}
	(\bibinfo{year}{2004}).
	
	\bibitem[{\citenamefont{Feynman}(1982)}]{feynman1982simulating}
	\bibinfo{author}{\bibfnamefont{R.~P.} \bibnamefont{Feynman}},
	\bibinfo{journal}{International Journal of Theoretical Physics}
	\textbf{\bibinfo{volume}{21}} (\bibinfo{year}{1982}).
	
	\bibitem[{\citenamefont{Buluta and Nori}(2009)}]{buluta2009quantum}
	\bibinfo{author}{\bibfnamefont{I.}~\bibnamefont{Buluta}} \bibnamefont{and}
	\bibinfo{author}{\bibfnamefont{F.}~\bibnamefont{Nori}},
	\bibinfo{journal}{Science} \textbf{\bibinfo{volume}{326}},
	\bibinfo{pages}{108} (\bibinfo{year}{2009}).
	
	\bibitem[{\citenamefont{Byrnes et~al.}(2007)\citenamefont{Byrnes, Recher, Kim,
			Utsunomiya, and Yamamoto}}]{byrnes2007quantum}
	\bibinfo{author}{\bibfnamefont{T.}~\bibnamefont{Byrnes}},
	\bibinfo{author}{\bibfnamefont{P.}~\bibnamefont{Recher}},
	\bibinfo{author}{\bibfnamefont{N.~Y.} \bibnamefont{Kim}},
	\bibinfo{author}{\bibfnamefont{S.}~\bibnamefont{Utsunomiya}},
	\bibnamefont{and} \bibinfo{author}{\bibfnamefont{Y.}~\bibnamefont{Yamamoto}},
	\bibinfo{journal}{Physical review letters} \textbf{\bibinfo{volume}{99}},
	\bibinfo{pages}{016405} (\bibinfo{year}{2007}).
	
	\bibitem[{\citenamefont{Georgescu et~al.}(2014)\citenamefont{Georgescu, Ashhab,
			and Nori}}]{georgescu2014quantum}
	\bibinfo{author}{\bibfnamefont{I.~M.} \bibnamefont{Georgescu}},
	\bibinfo{author}{\bibfnamefont{S.}~\bibnamefont{Ashhab}}, \bibnamefont{and}
	\bibinfo{author}{\bibfnamefont{F.}~\bibnamefont{Nori}},
	\bibinfo{journal}{Reviews of Modern Physics} \textbf{\bibinfo{volume}{86}},
	\bibinfo{pages}{153} (\bibinfo{year}{2014}).
	
	\bibitem[{\citenamefont{Cirac and Zoller}(2012)}]{cirac2012goals}
	\bibinfo{author}{\bibfnamefont{J.~I.} \bibnamefont{Cirac}} \bibnamefont{and}
	\bibinfo{author}{\bibfnamefont{P.}~\bibnamefont{Zoller}},
	\bibinfo{journal}{Nature physics} \textbf{\bibinfo{volume}{8}},
	\bibinfo{pages}{264} (\bibinfo{year}{2012}).
	
	\bibitem[{\citenamefont{Gerritsma et~al.}(2010)\citenamefont{Gerritsma,
			Kirchmair, Z{\"a}hringer, Solano, Blatt, and Roos}}]{gerritsma2010quantum}
	\bibinfo{author}{\bibfnamefont{R.}~\bibnamefont{Gerritsma}},
	\bibinfo{author}{\bibfnamefont{G.}~\bibnamefont{Kirchmair}},
	\bibinfo{author}{\bibfnamefont{F.}~\bibnamefont{Z{\"a}hringer}},
	\bibinfo{author}{\bibfnamefont{E.}~\bibnamefont{Solano}},
	\bibinfo{author}{\bibfnamefont{R.}~\bibnamefont{Blatt}}, \bibnamefont{and}
	\bibinfo{author}{\bibfnamefont{C.}~\bibnamefont{Roos}},
	\bibinfo{journal}{Nature} \textbf{\bibinfo{volume}{463}}, \bibinfo{pages}{68}
	(\bibinfo{year}{2010}).
	
	\bibitem[{\citenamefont{O’Malley et~al.}(2016)\citenamefont{O’Malley,
			Babbush, Kivlichan, Romero, McClean, Barends, Kelly, Roushan, Tranter, Ding
			et~al.}}]{o2016scalable}
	\bibinfo{author}{\bibfnamefont{P.~J.} \bibnamefont{O’Malley}},
	\bibinfo{author}{\bibfnamefont{R.}~\bibnamefont{Babbush}},
	\bibinfo{author}{\bibfnamefont{I.~D.} \bibnamefont{Kivlichan}},
	\bibinfo{author}{\bibfnamefont{J.}~\bibnamefont{Romero}},
	\bibinfo{author}{\bibfnamefont{J.~R.} \bibnamefont{McClean}},
	\bibinfo{author}{\bibfnamefont{R.}~\bibnamefont{Barends}},
	\bibinfo{author}{\bibfnamefont{J.}~\bibnamefont{Kelly}},
	\bibinfo{author}{\bibfnamefont{P.}~\bibnamefont{Roushan}},
	\bibinfo{author}{\bibfnamefont{A.}~\bibnamefont{Tranter}},
	\bibinfo{author}{\bibfnamefont{N.}~\bibnamefont{Ding}}, \bibnamefont{et~al.},
	\bibinfo{journal}{Physical Review X} \textbf{\bibinfo{volume}{6}},
	\bibinfo{pages}{031007} (\bibinfo{year}{2016}).
	
	\bibitem[{\citenamefont{Horikiri et~al.}(2016)\citenamefont{Horikiri,
			Yamaguchi, Kamide, Matsuo, Byrnes, Ishida, L{\"o}ffler, H{\"o}fling, Shikano,
			Ogawa et~al.}}]{horikiri2016high}
	\bibinfo{author}{\bibfnamefont{T.}~\bibnamefont{Horikiri}},
	\bibinfo{author}{\bibfnamefont{M.}~\bibnamefont{Yamaguchi}},
	\bibinfo{author}{\bibfnamefont{K.}~\bibnamefont{Kamide}},
	\bibinfo{author}{\bibfnamefont{Y.}~\bibnamefont{Matsuo}},
	\bibinfo{author}{\bibfnamefont{T.}~\bibnamefont{Byrnes}},
	\bibinfo{author}{\bibfnamefont{N.}~\bibnamefont{Ishida}},
	\bibinfo{author}{\bibfnamefont{A.}~\bibnamefont{L{\"o}ffler}},
	\bibinfo{author}{\bibfnamefont{S.}~\bibnamefont{H{\"o}fling}},
	\bibinfo{author}{\bibfnamefont{Y.}~\bibnamefont{Shikano}},
	\bibinfo{author}{\bibfnamefont{T.}~\bibnamefont{Ogawa}},
	\bibnamefont{et~al.}, \bibinfo{journal}{Scientific reports}
	\textbf{\bibinfo{volume}{6}}, \bibinfo{pages}{1} (\bibinfo{year}{2016}).
	
	\bibitem[{\citenamefont{Houck et~al.}(2012)\citenamefont{Houck, T{\"u}reci, and
			Koch}}]{houck2012chip}
	\bibinfo{author}{\bibfnamefont{A.~A.} \bibnamefont{Houck}},
	\bibinfo{author}{\bibfnamefont{H.~E.} \bibnamefont{T{\"u}reci}},
	\bibnamefont{and} \bibinfo{author}{\bibfnamefont{J.}~\bibnamefont{Koch}},
	\bibinfo{journal}{Nature Physics} \textbf{\bibinfo{volume}{8}},
	\bibinfo{pages}{292} (\bibinfo{year}{2012}).
	
	\bibitem[{\citenamefont{Byrnes and Ilo-Okeke}(2021)}]{byrnes2021quantum}
	\bibinfo{author}{\bibfnamefont{T.}~\bibnamefont{Byrnes}} \bibnamefont{and}
	\bibinfo{author}{\bibfnamefont{E.~O.} \bibnamefont{Ilo-Okeke}},
	\emph{\bibinfo{title}{Quantum atom optics: Theory and applications to quantum
			technology}} (\bibinfo{publisher}{Cambridge university press},
	\bibinfo{year}{2021}).
	
	\bibitem[{\citenamefont{Mohseni et~al.}(2022)\citenamefont{Mohseni, McMahon,
			and Byrnes}}]{mohseni2022ising}
	\bibinfo{author}{\bibfnamefont{N.}~\bibnamefont{Mohseni}},
	\bibinfo{author}{\bibfnamefont{P.~L.} \bibnamefont{McMahon}},
	\bibnamefont{and} \bibinfo{author}{\bibfnamefont{T.}~\bibnamefont{Byrnes}},
	\bibinfo{journal}{Nature Reviews Physics} \textbf{\bibinfo{volume}{4}},
	\bibinfo{pages}{363} (\bibinfo{year}{2022}).
	
	\bibitem[{\citenamefont{Lucas}(2014)}]{lucas2014ising}
	\bibinfo{author}{\bibfnamefont{A.}~\bibnamefont{Lucas}},
	\bibinfo{journal}{Frontiers in physics} \textbf{\bibinfo{volume}{2}},
	\bibinfo{pages}{5} (\bibinfo{year}{2014}).
	
	\bibitem[{\citenamefont{Tanahashi et~al.}(2019)\citenamefont{Tanahashi,
			Takayanagi, Motohashi, and Tanaka}}]{tanahashi2019application}
	\bibinfo{author}{\bibfnamefont{K.}~\bibnamefont{Tanahashi}},
	\bibinfo{author}{\bibfnamefont{S.}~\bibnamefont{Takayanagi}},
	\bibinfo{author}{\bibfnamefont{T.}~\bibnamefont{Motohashi}},
	\bibnamefont{and} \bibinfo{author}{\bibfnamefont{S.}~\bibnamefont{Tanaka}},
	\bibinfo{journal}{Journal of the Physical Society of Japan}
	\textbf{\bibinfo{volume}{88}}, \bibinfo{pages}{061010}
	(\bibinfo{year}{2019}).
	
	\bibitem[{\citenamefont{Smelyanskiy et~al.}(2012)\citenamefont{Smelyanskiy,
			Rieffel, Knysh, Williams, Johnson, Thom, Macready, and
			Pudenz}}]{smelyanskiy2012near}
	\bibinfo{author}{\bibfnamefont{V.~N.} \bibnamefont{Smelyanskiy}},
	\bibinfo{author}{\bibfnamefont{E.~G.} \bibnamefont{Rieffel}},
	\bibinfo{author}{\bibfnamefont{S.~I.} \bibnamefont{Knysh}},
	\bibinfo{author}{\bibfnamefont{C.~P.} \bibnamefont{Williams}},
	\bibinfo{author}{\bibfnamefont{M.~W.} \bibnamefont{Johnson}},
	\bibinfo{author}{\bibfnamefont{M.~C.} \bibnamefont{Thom}},
	\bibinfo{author}{\bibfnamefont{W.~G.} \bibnamefont{Macready}},
	\bibnamefont{and} \bibinfo{author}{\bibfnamefont{K.~L.}
		\bibnamefont{Pudenz}}, \bibinfo{journal}{arXiv preprint arXiv:1204.2821}
	(\bibinfo{year}{2012}).
	
	\bibitem[{\citenamefont{Hauke et~al.}(2020)\citenamefont{Hauke, Katzgraber,
			Lechner, Nishimori, and Oliver}}]{hauke2020perspectives}
	\bibinfo{author}{\bibfnamefont{P.}~\bibnamefont{Hauke}},
	\bibinfo{author}{\bibfnamefont{H.~G.} \bibnamefont{Katzgraber}},
	\bibinfo{author}{\bibfnamefont{W.}~\bibnamefont{Lechner}},
	\bibinfo{author}{\bibfnamefont{H.}~\bibnamefont{Nishimori}},
	\bibnamefont{and} \bibinfo{author}{\bibfnamefont{W.~D.}
		\bibnamefont{Oliver}}, \bibinfo{journal}{Reports on Progress in Physics}
	\textbf{\bibinfo{volume}{83}}, \bibinfo{pages}{054401}
	(\bibinfo{year}{2020}).
	
	\bibitem[{\citenamefont{Giovannetti et~al.}(2011)\citenamefont{Giovannetti,
			Lloyd, and Maccone}}]{giovannetti2011advances}
	\bibinfo{author}{\bibfnamefont{V.}~\bibnamefont{Giovannetti}},
	\bibinfo{author}{\bibfnamefont{S.}~\bibnamefont{Lloyd}}, \bibnamefont{and}
	\bibinfo{author}{\bibfnamefont{L.}~\bibnamefont{Maccone}},
	\bibinfo{journal}{Nature photonics} \textbf{\bibinfo{volume}{5}},
	\bibinfo{pages}{222} (\bibinfo{year}{2011}).
	
	\bibitem[{\citenamefont{T{\'o}th and Apellaniz}(2014)}]{toth2014quantum}
	\bibinfo{author}{\bibfnamefont{G.}~\bibnamefont{T{\'o}th}} \bibnamefont{and}
	\bibinfo{author}{\bibfnamefont{I.}~\bibnamefont{Apellaniz}},
	\bibinfo{journal}{Journal of Physics A: Mathematical and Theoretical}
	\textbf{\bibinfo{volume}{47}}, \bibinfo{pages}{424006}
	(\bibinfo{year}{2014}).
	
	\bibitem[{\citenamefont{You et~al.}(2017)\citenamefont{You, Adhikari, Chi,
			LaBorde, Matyas, Zhang, Su, Byrnes, Lu, Dowling
			et~al.}}]{you2017multiparameter}
	\bibinfo{author}{\bibfnamefont{C.}~\bibnamefont{You}},
	\bibinfo{author}{\bibfnamefont{S.}~\bibnamefont{Adhikari}},
	\bibinfo{author}{\bibfnamefont{Y.}~\bibnamefont{Chi}},
	\bibinfo{author}{\bibfnamefont{M.~L.} \bibnamefont{LaBorde}},
	\bibinfo{author}{\bibfnamefont{C.~T.} \bibnamefont{Matyas}},
	\bibinfo{author}{\bibfnamefont{C.}~\bibnamefont{Zhang}},
	\bibinfo{author}{\bibfnamefont{Z.}~\bibnamefont{Su}},
	\bibinfo{author}{\bibfnamefont{T.}~\bibnamefont{Byrnes}},
	\bibinfo{author}{\bibfnamefont{C.}~\bibnamefont{Lu}},
	\bibinfo{author}{\bibfnamefont{J.~P.} \bibnamefont{Dowling}},
	\bibnamefont{et~al.}, \bibinfo{journal}{Journal of Optics}
	\textbf{\bibinfo{volume}{19}}, \bibinfo{pages}{124002}
	(\bibinfo{year}{2017}).
	
	\bibitem[{\citenamefont{Raussendorf and Briegel}(2001)}]{raussendorf2001one}
	\bibinfo{author}{\bibfnamefont{R.}~\bibnamefont{Raussendorf}} \bibnamefont{and}
	\bibinfo{author}{\bibfnamefont{H.~J.} \bibnamefont{Briegel}},
	\bibinfo{journal}{Physical Review Letters} \textbf{\bibinfo{volume}{86}},
	\bibinfo{pages}{5188} (\bibinfo{year}{2001}).
	
	\bibitem[{\citenamefont{Nayak et~al.}(2008)\citenamefont{Nayak, Simon, Stern,
			Freedman, and Sarma}}]{nayak2008non}
	\bibinfo{author}{\bibfnamefont{C.}~\bibnamefont{Nayak}},
	\bibinfo{author}{\bibfnamefont{S.~H.} \bibnamefont{Simon}},
	\bibinfo{author}{\bibfnamefont{A.}~\bibnamefont{Stern}},
	\bibinfo{author}{\bibfnamefont{M.}~\bibnamefont{Freedman}}, \bibnamefont{and}
	\bibinfo{author}{\bibfnamefont{S.~D.} \bibnamefont{Sarma}},
	\bibinfo{journal}{Reviews of Modern Physics} \textbf{\bibinfo{volume}{80}},
	\bibinfo{pages}{1083} (\bibinfo{year}{2008}).
	
	\bibitem[{\citenamefont{Abdelrahman et~al.}(2014)\citenamefont{Abdelrahman,
			Mukai, H{\"a}ffner, and Byrnes}}]{abdelrahman2014coherent}
	\bibinfo{author}{\bibfnamefont{A.}~\bibnamefont{Abdelrahman}},
	\bibinfo{author}{\bibfnamefont{T.}~\bibnamefont{Mukai}},
	\bibinfo{author}{\bibfnamefont{H.}~\bibnamefont{H{\"a}ffner}},
	\bibnamefont{and} \bibinfo{author}{\bibfnamefont{T.}~\bibnamefont{Byrnes}},
	\bibinfo{journal}{Optics express} \textbf{\bibinfo{volume}{22}},
	\bibinfo{pages}{3501} (\bibinfo{year}{2014}).
	
	\bibitem[{\citenamefont{Tame et~al.}(2006)\citenamefont{Tame, Paternostro, Kim,
			and Vedral}}]{tame2006natural}
	\bibinfo{author}{\bibfnamefont{M.}~\bibnamefont{Tame}},
	\bibinfo{author}{\bibfnamefont{M.}~\bibnamefont{Paternostro}},
	\bibinfo{author}{\bibfnamefont{M.}~\bibnamefont{Kim}}, \bibnamefont{and}
	\bibinfo{author}{\bibfnamefont{V.}~\bibnamefont{Vedral}},
	\bibinfo{journal}{Physical Review A} \textbf{\bibinfo{volume}{73}},
	\bibinfo{pages}{022309} (\bibinfo{year}{2006}).
	
	\bibitem[{\citenamefont{Bartlett and Rudolph}(2006)}]{bartlett2006simple}
	\bibinfo{author}{\bibfnamefont{S.~D.} \bibnamefont{Bartlett}} \bibnamefont{and}
	\bibinfo{author}{\bibfnamefont{T.}~\bibnamefont{Rudolph}},
	\bibinfo{journal}{Physical Review A} \textbf{\bibinfo{volume}{74}},
	\bibinfo{pages}{040302} (\bibinfo{year}{2006}).
	
	\bibitem[{\citenamefont{Van~den Nest et~al.}(2008)\citenamefont{Van~den Nest,
			Luttmer, D{\"u}r, and Briegel}}]{van2008graph}
	\bibinfo{author}{\bibfnamefont{M.}~\bibnamefont{Van~den Nest}},
	\bibinfo{author}{\bibfnamefont{K.}~\bibnamefont{Luttmer}},
	\bibinfo{author}{\bibfnamefont{W.}~\bibnamefont{D{\"u}r}}, \bibnamefont{and}
	\bibinfo{author}{\bibfnamefont{H.}~\bibnamefont{Briegel}},
	\bibinfo{journal}{Physical Review A} \textbf{\bibinfo{volume}{77}},
	\bibinfo{pages}{012301} (\bibinfo{year}{2008}).
	
	\bibitem[{\citenamefont{Kyaw et~al.}(2014)\citenamefont{Kyaw, Li, and
			Kwek}}]{kyaw2014measurement}
	\bibinfo{author}{\bibfnamefont{T.~H.} \bibnamefont{Kyaw}},
	\bibinfo{author}{\bibfnamefont{Y.}~\bibnamefont{Li}}, \bibnamefont{and}
	\bibinfo{author}{\bibfnamefont{L.-C.} \bibnamefont{Kwek}},
	\bibinfo{journal}{Physical Review Letters} \textbf{\bibinfo{volume}{113}},
	\bibinfo{pages}{180501} (\bibinfo{year}{2014}).
	
	\bibitem[{\citenamefont{Jones et~al.}(2019)\citenamefont{Jones, Endo, McArdle,
			Yuan, and Benjamin}}]{jones2019variational}
	\bibinfo{author}{\bibfnamefont{T.}~\bibnamefont{Jones}},
	\bibinfo{author}{\bibfnamefont{S.}~\bibnamefont{Endo}},
	\bibinfo{author}{\bibfnamefont{S.}~\bibnamefont{McArdle}},
	\bibinfo{author}{\bibfnamefont{X.}~\bibnamefont{Yuan}}, \bibnamefont{and}
	\bibinfo{author}{\bibfnamefont{S.~C.} \bibnamefont{Benjamin}},
	\bibinfo{journal}{Physical Review A} \textbf{\bibinfo{volume}{99}},
	\bibinfo{pages}{062304} (\bibinfo{year}{2019}).
	
	\bibitem[{\citenamefont{Endo et~al.}(2020)\citenamefont{Endo, Sun, Li,
			Benjamin, and Yuan}}]{endo2020variational}
	\bibinfo{author}{\bibfnamefont{S.}~\bibnamefont{Endo}},
	\bibinfo{author}{\bibfnamefont{J.}~\bibnamefont{Sun}},
	\bibinfo{author}{\bibfnamefont{Y.}~\bibnamefont{Li}},
	\bibinfo{author}{\bibfnamefont{S.~C.} \bibnamefont{Benjamin}},
	\bibnamefont{and} \bibinfo{author}{\bibfnamefont{X.}~\bibnamefont{Yuan}},
	\bibinfo{journal}{Physical Review Letters} \textbf{\bibinfo{volume}{125}},
	\bibinfo{pages}{010501} (\bibinfo{year}{2020}).
	
	\bibitem[{\citenamefont{Yuan et~al.}(2019)\citenamefont{Yuan, Endo, Zhao, Li,
			and Benjamin}}]{yuan2019theory}
	\bibinfo{author}{\bibfnamefont{X.}~\bibnamefont{Yuan}},
	\bibinfo{author}{\bibfnamefont{S.}~\bibnamefont{Endo}},
	\bibinfo{author}{\bibfnamefont{Q.}~\bibnamefont{Zhao}},
	\bibinfo{author}{\bibfnamefont{Y.}~\bibnamefont{Li}}, \bibnamefont{and}
	\bibinfo{author}{\bibfnamefont{S.~C.} \bibnamefont{Benjamin}},
	\bibinfo{journal}{Quantum} \textbf{\bibinfo{volume}{3}}, \bibinfo{pages}{191}
	(\bibinfo{year}{2019}).
	
	\bibitem[{\citenamefont{Motta et~al.}(2020)\citenamefont{Motta, Sun, Tan,
			O’Rourke, Ye, Minnich, Brand{\~a}o, and Chan}}]{motta2020determining}
	\bibinfo{author}{\bibfnamefont{M.}~\bibnamefont{Motta}},
	\bibinfo{author}{\bibfnamefont{C.}~\bibnamefont{Sun}},
	\bibinfo{author}{\bibfnamefont{A.~T.} \bibnamefont{Tan}},
	\bibinfo{author}{\bibfnamefont{M.~J.} \bibnamefont{O’Rourke}},
	\bibinfo{author}{\bibfnamefont{E.}~\bibnamefont{Ye}},
	\bibinfo{author}{\bibfnamefont{A.~J.} \bibnamefont{Minnich}},
	\bibinfo{author}{\bibfnamefont{F.~G.} \bibnamefont{Brand{\~a}o}},
	\bibnamefont{and} \bibinfo{author}{\bibfnamefont{G.~K.-L.}
		\bibnamefont{Chan}}, \bibinfo{journal}{Nature Physics}
	\textbf{\bibinfo{volume}{16}}, \bibinfo{pages}{205} (\bibinfo{year}{2020}).
	
	\bibitem[{\citenamefont{Yeter-Aydeniz et~al.}(2020)\citenamefont{Yeter-Aydeniz,
			Pooser, and Siopsis}}]{yeter2020practical}
	\bibinfo{author}{\bibfnamefont{K.}~\bibnamefont{Yeter-Aydeniz}},
	\bibinfo{author}{\bibfnamefont{R.~C.} \bibnamefont{Pooser}},
	\bibnamefont{and} \bibinfo{author}{\bibfnamefont{G.}~\bibnamefont{Siopsis}},
	\bibinfo{journal}{npj Quantum Information} \textbf{\bibinfo{volume}{6}},
	\bibinfo{pages}{1} (\bibinfo{year}{2020}).
	
	\bibitem[{\citenamefont{Gomes et~al.}(2020)\citenamefont{Gomes, Zhang,
			Berthusen, Wang, Ho, Orth, and Yao}}]{gomes2020efficient}
	\bibinfo{author}{\bibfnamefont{N.}~\bibnamefont{Gomes}},
	\bibinfo{author}{\bibfnamefont{F.}~\bibnamefont{Zhang}},
	\bibinfo{author}{\bibfnamefont{N.~F.} \bibnamefont{Berthusen}},
	\bibinfo{author}{\bibfnamefont{C.-Z.} \bibnamefont{Wang}},
	\bibinfo{author}{\bibfnamefont{K.-M.} \bibnamefont{Ho}},
	\bibinfo{author}{\bibfnamefont{P.~P.} \bibnamefont{Orth}}, \bibnamefont{and}
	\bibinfo{author}{\bibfnamefont{Y.}~\bibnamefont{Yao}},
	\bibinfo{journal}{Journal of Chemical Theory and Computation}
	\textbf{\bibinfo{volume}{16}}, \bibinfo{pages}{6256} (\bibinfo{year}{2020}).
	
	\bibitem[{\citenamefont{Tan}(2020)}]{tan2020fast}
	\bibinfo{author}{\bibfnamefont{K.~C.} \bibnamefont{Tan}},
	\bibinfo{journal}{arXiv preprint arXiv:2009.12239}  (\bibinfo{year}{2020}).
	
	\bibitem[{\citenamefont{Kamakari et~al.}(2022)\citenamefont{Kamakari, Sun,
			Motta, and Minnich}}]{kamakari2022digital}
	\bibinfo{author}{\bibfnamefont{H.}~\bibnamefont{Kamakari}},
	\bibinfo{author}{\bibfnamefont{S.-N.} \bibnamefont{Sun}},
	\bibinfo{author}{\bibfnamefont{M.}~\bibnamefont{Motta}}, \bibnamefont{and}
	\bibinfo{author}{\bibfnamefont{A.~J.} \bibnamefont{Minnich}},
	\bibinfo{journal}{PRX Quantum} \textbf{\bibinfo{volume}{3}},
	\bibinfo{pages}{010320} (\bibinfo{year}{2022}).
	
	\bibitem[{\citenamefont{Cao et~al.}(2022)\citenamefont{Cao, An, Hou, Zhou, and
			Zeng}}]{cao2022quantum}
	\bibinfo{author}{\bibfnamefont{C.}~\bibnamefont{Cao}},
	\bibinfo{author}{\bibfnamefont{Z.}~\bibnamefont{An}},
	\bibinfo{author}{\bibfnamefont{S.-Y.} \bibnamefont{Hou}},
	\bibinfo{author}{\bibfnamefont{D.}~\bibnamefont{Zhou}}, \bibnamefont{and}
	\bibinfo{author}{\bibfnamefont{B.}~\bibnamefont{Zeng}},
	\bibinfo{journal}{Communications Physics} \textbf{\bibinfo{volume}{5}},
	\bibinfo{pages}{1} (\bibinfo{year}{2022}).
	
	\bibitem[{\citenamefont{Nishi et~al.}(2021)\citenamefont{Nishi, Kosugi, and
			Matsushita}}]{nishi2021implementation}
	\bibinfo{author}{\bibfnamefont{H.}~\bibnamefont{Nishi}},
	\bibinfo{author}{\bibfnamefont{T.}~\bibnamefont{Kosugi}}, \bibnamefont{and}
	\bibinfo{author}{\bibfnamefont{Y.-i.} \bibnamefont{Matsushita}},
	\bibinfo{journal}{npj Quantum Information} \textbf{\bibinfo{volume}{7}},
	\bibinfo{pages}{1} (\bibinfo{year}{2021}).
	
	\bibitem[{\citenamefont{Williams}(2004)}]{williams2004probabilistic}
	\bibinfo{author}{\bibfnamefont{C.~P.} \bibnamefont{Williams}}, in
	\emph{\bibinfo{booktitle}{Quantum Information and Computation II}}
	(\bibinfo{organization}{International Society for Optics and Photonics},
	\bibinfo{year}{2004}), vol. \bibinfo{volume}{5436}, pp.
	\bibinfo{pages}{297--306}.
	
	\bibitem[{\citenamefont{Liu et~al.}(2021)\citenamefont{Liu, Liu, and
			Fan}}]{liu2021probabilistic}
	\bibinfo{author}{\bibfnamefont{T.}~\bibnamefont{Liu}},
	\bibinfo{author}{\bibfnamefont{J.-G.} \bibnamefont{Liu}}, \bibnamefont{and}
	\bibinfo{author}{\bibfnamefont{H.}~\bibnamefont{Fan}},
	\bibinfo{journal}{Quantum Information Processing}
	\textbf{\bibinfo{volume}{20}}, \bibinfo{pages}{1} (\bibinfo{year}{2021}).
	
	\bibitem[{\citenamefont{Gingrich and Williams}(2004)}]{gingrich2004non}
	\bibinfo{author}{\bibfnamefont{R.~M.} \bibnamefont{Gingrich}} \bibnamefont{and}
	\bibinfo{author}{\bibfnamefont{C.~P.} \bibnamefont{Williams}}
	(\bibinfo{year}{2004}).
	
	\bibitem[{\citenamefont{Grover}(1997)}]{PhysRevLett.79.325}
	\bibinfo{author}{\bibfnamefont{L.~K.} \bibnamefont{Grover}},
	\bibinfo{journal}{Phys. Rev. Lett.} \textbf{\bibinfo{volume}{79}},
	\bibinfo{pages}{325} (\bibinfo{year}{1997}).
	
	\bibitem[{\citenamefont{Kosugi et~al.}(2021)\citenamefont{Kosugi, Nishiya, and
			Matsushita}}]{kosugi2021probabilistic}
	\bibinfo{author}{\bibfnamefont{T.}~\bibnamefont{Kosugi}},
	\bibinfo{author}{\bibfnamefont{Y.}~\bibnamefont{Nishiya}}, \bibnamefont{and}
	\bibinfo{author}{\bibfnamefont{Y.-i.} \bibnamefont{Matsushita}},
	\bibinfo{journal}{arXiv preprint arXiv:2111.12471}  (\bibinfo{year}{2021}).
	
	\bibitem[{\citenamefont{Shingu et~al.}(2021)\citenamefont{Shingu, Seki, Watabe,
			Endo, Matsuzaki, Kawabata, Nikuni, and Hakoshima}}]{shingu2021boltzmann}
	\bibinfo{author}{\bibfnamefont{Y.}~\bibnamefont{Shingu}},
	\bibinfo{author}{\bibfnamefont{Y.}~\bibnamefont{Seki}},
	\bibinfo{author}{\bibfnamefont{S.}~\bibnamefont{Watabe}},
	\bibinfo{author}{\bibfnamefont{S.}~\bibnamefont{Endo}},
	\bibinfo{author}{\bibfnamefont{Y.}~\bibnamefont{Matsuzaki}},
	\bibinfo{author}{\bibfnamefont{S.}~\bibnamefont{Kawabata}},
	\bibinfo{author}{\bibfnamefont{T.}~\bibnamefont{Nikuni}}, \bibnamefont{and}
	\bibinfo{author}{\bibfnamefont{H.}~\bibnamefont{Hakoshima}},
	\bibinfo{journal}{Physical Review A} \textbf{\bibinfo{volume}{104}},
	\bibinfo{pages}{032413} (\bibinfo{year}{2021}).
	
	\bibitem[{\citenamefont{Yeter-Aydeniz et~al.}(2021)\citenamefont{Yeter-Aydeniz,
			Siopsis, and Pooser}}]{yeter2021scattering}
	\bibinfo{author}{\bibfnamefont{K.}~\bibnamefont{Yeter-Aydeniz}},
	\bibinfo{author}{\bibfnamefont{G.}~\bibnamefont{Siopsis}}, \bibnamefont{and}
	\bibinfo{author}{\bibfnamefont{R.~C.} \bibnamefont{Pooser}},
	\bibinfo{journal}{New Journal of Physics} \textbf{\bibinfo{volume}{23}},
	\bibinfo{pages}{043033} (\bibinfo{year}{2021}).
	
	\bibitem[{\citenamefont{Zeng et~al.}(2021)\citenamefont{Zeng, Cao, Zhang, Xu,
			and Zeng}}]{zeng2021variational}
	\bibinfo{author}{\bibfnamefont{J.}~\bibnamefont{Zeng}},
	\bibinfo{author}{\bibfnamefont{C.}~\bibnamefont{Cao}},
	\bibinfo{author}{\bibfnamefont{C.}~\bibnamefont{Zhang}},
	\bibinfo{author}{\bibfnamefont{P.}~\bibnamefont{Xu}}, \bibnamefont{and}
	\bibinfo{author}{\bibfnamefont{B.}~\bibnamefont{Zeng}},
	\bibinfo{journal}{Quantum Science and Technology}
	\textbf{\bibinfo{volume}{6}}, \bibinfo{pages}{045009} (\bibinfo{year}{2021}).
	
	\bibitem[{\citenamefont{Knill et~al.}(2001)\citenamefont{Knill, Laflamme, and
			Milburn}}]{knill2001scheme}
	\bibinfo{author}{\bibfnamefont{E.}~\bibnamefont{Knill}},
	\bibinfo{author}{\bibfnamefont{R.}~\bibnamefont{Laflamme}}, \bibnamefont{and}
	\bibinfo{author}{\bibfnamefont{G.~J.} \bibnamefont{Milburn}},
	\bibinfo{journal}{nature} \textbf{\bibinfo{volume}{409}}, \bibinfo{pages}{46}
	(\bibinfo{year}{2001}).
	
	\bibitem[{\citenamefont{Steffen et~al.}(2013)\citenamefont{Steffen, Salathe,
			Oppliger, Kurpiers, Baur, Lang, Eichler, Puebla-Hellmann, Fedorov, and
			Wallraff}}]{steffen2013deterministic}
	\bibinfo{author}{\bibfnamefont{L.}~\bibnamefont{Steffen}},
	\bibinfo{author}{\bibfnamefont{Y.}~\bibnamefont{Salathe}},
	\bibinfo{author}{\bibfnamefont{M.}~\bibnamefont{Oppliger}},
	\bibinfo{author}{\bibfnamefont{P.}~\bibnamefont{Kurpiers}},
	\bibinfo{author}{\bibfnamefont{M.}~\bibnamefont{Baur}},
	\bibinfo{author}{\bibfnamefont{C.}~\bibnamefont{Lang}},
	\bibinfo{author}{\bibfnamefont{C.}~\bibnamefont{Eichler}},
	\bibinfo{author}{\bibfnamefont{G.}~\bibnamefont{Puebla-Hellmann}},
	\bibinfo{author}{\bibfnamefont{A.}~\bibnamefont{Fedorov}}, \bibnamefont{and}
	\bibinfo{author}{\bibfnamefont{A.}~\bibnamefont{Wallraff}},
	\bibinfo{journal}{Nature} \textbf{\bibinfo{volume}{500}},
	\bibinfo{pages}{319} (\bibinfo{year}{2013}).
	
	\bibitem[{\citenamefont{Ma et~al.}(2012)\citenamefont{Ma, Herbst, Scheidl,
			Wang, Kropatschek, Naylor, Wittmann, Mech, Kofler, Anisimova
			et~al.}}]{ma2012quantum}
	\bibinfo{author}{\bibfnamefont{X.-S.} \bibnamefont{Ma}},
	\bibinfo{author}{\bibfnamefont{T.}~\bibnamefont{Herbst}},
	\bibinfo{author}{\bibfnamefont{T.}~\bibnamefont{Scheidl}},
	\bibinfo{author}{\bibfnamefont{D.}~\bibnamefont{Wang}},
	\bibinfo{author}{\bibfnamefont{S.}~\bibnamefont{Kropatschek}},
	\bibinfo{author}{\bibfnamefont{W.}~\bibnamefont{Naylor}},
	\bibinfo{author}{\bibfnamefont{B.}~\bibnamefont{Wittmann}},
	\bibinfo{author}{\bibfnamefont{A.}~\bibnamefont{Mech}},
	\bibinfo{author}{\bibfnamefont{J.}~\bibnamefont{Kofler}},
	\bibinfo{author}{\bibfnamefont{E.}~\bibnamefont{Anisimova}},
	\bibnamefont{et~al.}, \bibinfo{journal}{Nature}
	\textbf{\bibinfo{volume}{489}}, \bibinfo{pages}{269} (\bibinfo{year}{2012}).
	
	\bibitem[{\citenamefont{Ilo-Okeke et~al.}(2018)\citenamefont{Ilo-Okeke,
			Tessler, Dowling, and Byrnes}}]{ilo2018remote}
	\bibinfo{author}{\bibfnamefont{E.~O.} \bibnamefont{Ilo-Okeke}},
	\bibinfo{author}{\bibfnamefont{L.}~\bibnamefont{Tessler}},
	\bibinfo{author}{\bibfnamefont{J.~P.} \bibnamefont{Dowling}},
	\bibnamefont{and} \bibinfo{author}{\bibfnamefont{T.}~\bibnamefont{Byrnes}},
	\bibinfo{journal}{npj Quantum Information} \textbf{\bibinfo{volume}{4}},
	\bibinfo{pages}{1} (\bibinfo{year}{2018}).
	
	\bibitem[{\citenamefont{Ilo-Okeke and Byrnes}(2014)}]{ilo2014theory}
	\bibinfo{author}{\bibfnamefont{E.~O.} \bibnamefont{Ilo-Okeke}}
	\bibnamefont{and} \bibinfo{author}{\bibfnamefont{T.}~\bibnamefont{Byrnes}},
	\bibinfo{journal}{Physical review letters} \textbf{\bibinfo{volume}{112}},
	\bibinfo{pages}{233602} (\bibinfo{year}{2014}).
	
	\bibitem[{\citenamefont{Ilo-Okeke and Byrnes}(2016)}]{ilo2016information}
	\bibinfo{author}{\bibfnamefont{E.~O.} \bibnamefont{Ilo-Okeke}}
	\bibnamefont{and} \bibinfo{author}{\bibfnamefont{T.}~\bibnamefont{Byrnes}},
	\bibinfo{journal}{Physical Review A} \textbf{\bibinfo{volume}{94}},
	\bibinfo{pages}{013617} (\bibinfo{year}{2016}).
	
	\bibitem[{\citenamefont{Suzuki}(1993)}]{suzuki1993improved}
	\bibinfo{author}{\bibfnamefont{M.}~\bibnamefont{Suzuki}},
	\bibinfo{journal}{Physics Letters A} \textbf{\bibinfo{volume}{180}},
	\bibinfo{pages}{232} (\bibinfo{year}{1993}).
	
	\bibitem[{\citenamefont{Kapit et~al.}(2012)\citenamefont{Kapit, Ginsparg, and
			Mueller}}]{kapit2012non}
	\bibinfo{author}{\bibfnamefont{E.}~\bibnamefont{Kapit}},
	\bibinfo{author}{\bibfnamefont{P.}~\bibnamefont{Ginsparg}}, \bibnamefont{and}
	\bibinfo{author}{\bibfnamefont{E.}~\bibnamefont{Mueller}},
	\bibinfo{journal}{Physical Review Letters} \textbf{\bibinfo{volume}{108}},
	\bibinfo{pages}{066802} (\bibinfo{year}{2012}).
	
	\bibitem[{\citenamefont{Ochoa et~al.}(2018)\citenamefont{Ochoa, Belzig, and
			Nitzan}}]{ochoa2018simultaneous}
	\bibinfo{author}{\bibfnamefont{M.~A.} \bibnamefont{Ochoa}},
	\bibinfo{author}{\bibfnamefont{W.}~\bibnamefont{Belzig}}, \bibnamefont{and}
	\bibinfo{author}{\bibfnamefont{A.}~\bibnamefont{Nitzan}},
	\bibinfo{journal}{Scientific reports} \textbf{\bibinfo{volume}{8}},
	\bibinfo{pages}{1} (\bibinfo{year}{2018}).
	
	\bibitem[{\citenamefont{Ilo-Okeke et~al.}(2023)\citenamefont{Ilo-Okeke, Chen,
			Li, Anusionwu, Ivannikov, and Byrnes}}]{ilo2023measurement}
	\bibinfo{author}{\bibfnamefont{E.~O.} \bibnamefont{Ilo-Okeke}},
	\bibinfo{author}{\bibfnamefont{P.}~\bibnamefont{Chen}},
	\bibinfo{author}{\bibfnamefont{S.}~\bibnamefont{Li}},
	\bibinfo{author}{\bibfnamefont{B.~C.} \bibnamefont{Anusionwu}},
	\bibinfo{author}{\bibfnamefont{V.}~\bibnamefont{Ivannikov}},
	\bibnamefont{and} \bibinfo{author}{\bibfnamefont{T.}~\bibnamefont{Byrnes}},
	\bibinfo{journal}{AVS Quantum Science} \textbf{\bibinfo{volume}{5}}
	(\bibinfo{year}{2023}).
	
	\bibitem[{\citenamefont{Chaudhary et~al.}(2023)\citenamefont{Chaudhary,
			Ilo-Okeke, Ivannikov, and Byrnes}}]{chaudhary2023macroscopic}
	\bibinfo{author}{\bibfnamefont{M.}~\bibnamefont{Chaudhary}},
	\bibinfo{author}{\bibfnamefont{E.~O.} \bibnamefont{Ilo-Okeke}},
	\bibinfo{author}{\bibfnamefont{V.}~\bibnamefont{Ivannikov}},
	\bibnamefont{and} \bibinfo{author}{\bibfnamefont{T.}~\bibnamefont{Byrnes}},
	\bibinfo{journal}{arXiv preprint arXiv:2302.07526}  (\bibinfo{year}{2023}).
	
	\bibitem[{\citenamefont{Kondappan et~al.}(2023)\citenamefont{Kondappan,
			Chaudhary, Ilo-Okeke, Ivannikov, and Byrnes}}]{kondappan2023imaginary}
	\bibinfo{author}{\bibfnamefont{M.}~\bibnamefont{Kondappan}},
	\bibinfo{author}{\bibfnamefont{M.}~\bibnamefont{Chaudhary}},
	\bibinfo{author}{\bibfnamefont{E.~O.} \bibnamefont{Ilo-Okeke}},
	\bibinfo{author}{\bibfnamefont{V.}~\bibnamefont{Ivannikov}},
	\bibnamefont{and} \bibinfo{author}{\bibfnamefont{T.}~\bibnamefont{Byrnes}},
	\bibinfo{journal}{Physical Review A} \textbf{\bibinfo{volume}{107}},
	\bibinfo{pages}{042616} (\bibinfo{year}{2023}).
	
\end{thebibliography}
\end{document}